\documentclass{aa}  
\usepackage{graphicx}
\usepackage{txfonts}
\usepackage[colorlinks=true,citecolor=blue]{hyperref}

\usepackage{array}  
\usepackage{booktabs} 
\usepackage{multirow} 
\usepackage{placeins}          
\usepackage{float}

\newcommand{\ii}{{{\rm i}}}

\usepackage{xcolor}
\usepackage{amssymb}
\usepackage{amsmath} 
\usepackage{mathrsfs} 
\usepackage{stmaryrd}
\usepackage{dsfont}
\usepackage[T1]{fontenc}
\usepackage{txfonts}
\usepackage{bm}
\usepackage{mathtools}
\usepackage{empheq}
\usepackage{comment}
\usepackage{natbib,twoopt}
\bibpunct{(}{)}{;}{a}{}{,}            
\makeatletter
  \newcommandtwoopt{\citeads}[3][][]{\href{http://adsabs.harvard.edu/abs/#3}
    {\def\hyper@linkstart##1##2{}
     \let\hyper@linkend\@empty\citealp[#1][#2]{#3}}}
  \newcommandtwoopt{\citepads}[3][][]{\href{http://adsabs.harvard.edu/abs/#3}
    {\def\hyper@linkstart##1##2{}
     \let\hyper@linkend\@empty\citep[#1][#2]{#3}}}
  \newcommandtwoopt{\citetads}[3][][]{\href{http://adsabs.harvard.edu/abs/#3}
    {\def\hyper@linkstart##1##2{}
     \let\hyper@linkend\@empty\citet[#1][#2]{#3}}}
  \newcommandtwoopt{\citeyearads}[3][][]
    {\href{http://adsabs.harvard.edu/abs/#3}
    {\def\hyper@linkstart##1##2{}
     \let\hyper@linkend\@empty\citeyear[#1][#2]{#3}}}
    \renewcommand*\aa@pageof{, page \thepage{} of \pageref*{LastPage}}
\makeatother

\begin{document}

   \title{ 
   Effects of the radiative interior on solar inertial modes}
\author{Suprabha Mukhopadhyay \inst{1}
\and Yuto Bekki \inst{1}
\and Xiaojue Zhu \inst{1} \thanks{Corresponding author: \email{\href{mailto:zhux@mps.mpg.de}{zhux@mps.mpg.de}}}
\and Laurent Gizon \inst{1,2} \thanks{Corresponding author: \email{\href{mailto:gizon@mps.mpg.de}{gizon@mps.mpg.de}}}
} 
\institute{Max-Planck-Institut f{\"u}r Sonnensystemforschung,
              Justus-von-Liebig-Weg 3, 37077 G{\"o}ttingen, Germany
         \and
         Institut f{\"u}r Astrophysik und Geophysik, Georg-August-Universit{\"a}t G{\"o}ttingen,
         Friedrich-Hund-Platz 1, 37077 G{\"o}ttingen, Germany}

   \date{Received / accepted }

\abstract
{
Solar inertial modes are believed to play important diagnostic and dynamical roles in the Sun's differentially rotating convection zone. However, the coupling of these modes to the radiative interior has not yet been discussed.
}
{
We aim to understand the dependence of the modes on the uniformly rotating sub-adiabatic region below the convection zone and determine whether this leads to measurable changes at the surface.
}
{
We used the \texttt{Dedalus} code to compute linear eigenmodes in the inertial frequency range in a setup that includes both the convection zone and the radiative interior down to $0.5R_\odot$. We imposed free-surface boundary conditions at both radial boundaries. For comparison, we also computed the eigenmodes in a setup restricted to the convection zone. 
}
{
We find that including the radiative zone only slightly modifies the frequencies and surface eigenfunctions, except for some modes with significant radial motion (high-frequency retrograde and prograde columnar modes). On the other hand, most modes penetrate significantly into the overshooting layer below the convection zone. This reduces their growth rates and distorts their eigenfunctions near the base of the convection zone. Furthermore, the uniformly rotating sub-adiabatic radiative zone supports oscillations due to Rossby modes of all possible spherical harmonics and radial nodes. In particular, when the nearest inertial mode in frequency space lies within around 10 nHz and shares the same north--south symmetry, these Rossby modes evolve into mixed modes characterized by significant motions within both the radiative and convection zones. However, such mixed modes have a high mode mass in the radiative interior and thus will be difficult to excite stochastically via convection.
}
{}


   \maketitle

\section{Introduction}
\begin{figure*}
    \centering   \includegraphics[width=0.9\linewidth]{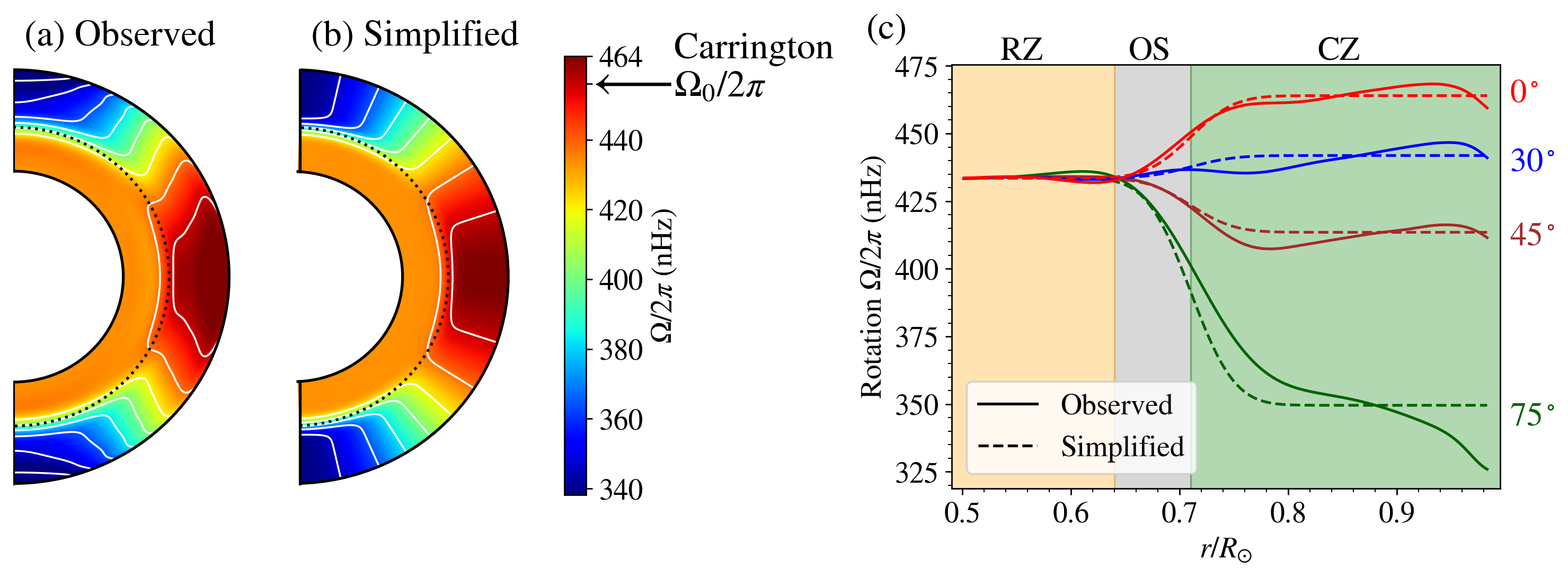}
    \caption{Profiles of differential rotation used in this study.
    (a) Observed 2D profile in a meridional plane, taken from \citet{Larson2018SoPh}.
    (b)~Simplified differential rotation profile, given by Eq.~\eqref{eq:simpleDR}, used to model the observed differential rotation. The dotted black lines indicate the base of the CZ at $r=0.71R_{\odot}$. The Carrington rotation rate, $\Omega_0/2\pi=456~\rm nHz,$ is marked on the colour bar. 
    (c) Cuts of observed (solid) and simplified (dashed) differential rotation profiles at fixed latitudes: $0^\circ$ (red), $30^\circ$ (blue), $45^\circ$ (brown), and $75^\circ$ (dark green). The green-shaded area denotes the CZ, the grey-shaded region the OS, and the orange-shaded region the RZ.
    }
    \label{fig:1}
\end{figure*}

Inertial modes are global oscillations in a rotating fluid whose restoring force is the Coriolis force. 
Different classes of inertial modes have been identified in the solar near-surface flows using various observational datasets and techniques (see \citealt{Gizon2024IAU} for a review). 
These include equatorial Rossby modes \citep{Loptien2018NatAs}, critical-layer and high-latitude modes \citep{Gizon2021A&A}, and so-called high-frequency retrograde (HFR) modes \citep{Hanson2022NatAs}.
All observed modes propagate retrograde with respect to the fastest local rotation rate, i.e. the surface equatorial rotation rate (or the Carrington rotation rate).
The solar inertial modes are found to be very sensitive to the solar differential rotation in the convection zone (CZ; Fig.~\ref{fig:1}a), as well as to several unknown quantities in the deep interior, such as turbulent viscosity and super-adiabaticity \citep[e.g.][]{Gizon2021A&A, Bekki2022A&A, Fournier2022A&A, Hanson2024PhFl}. Diagnosing these physical quantities using inertial modes may improve our understanding of the global-scale flow dynamics. 
Apart from their diagnostic role, the solar inertial modes may also play a significant dynamical role by providing non-linear feedback to regulate the Sun's differential rotation \citep{Bekki2024SciA}.

Linear models in 2D and 3D have helped identify the inertial modes and understand their physics \citep[e.g.][]{Gizon2020A&A, Bekki2022A&A, Fournier2022A&A, Bhattacharya2023ApJs, Mukhopadhyay2025A&A}. 
So far, most theoretical studies of inertial modes have been conducted only in the CZ and with impenetrable radial boundary conditions. 
It remains unclear how the radiative interior affects the properties of solar inertial modes in the convective envelope.

\citet{Blume2024ApJ} recently presented a non-linear simulation including the radiative interior and showed that both sectoral and tesseral Rossby modes are ubiquitously found in the radiative zone (RZ).
It is implied that these toroidal Rossby modes are likely present in a wave cavity different from the one in the CZ. 
\citet{Matilsky2022ApJ} further proposed that these Rossby modes might play a role in the confinement of the tachocline or in generating the local dynamo in the RZ. 
Neither study, however, reported the spatial eigenfunctions of these Rossby modes, and the mode coupling between the radiative and convective zones is still not well understood.

The presence of a convectively stable stratification allows for the existence of (gravito-)inertial modes whose restoring forces are both buoyancy and Coriolis forces. 
Gravito-inertial modes are important for inferring rotation profiles in $\gamma$ Doradus stars, which have convective cores and radiative envelopes \citep[e.g.][]{Ouazzani2020A&A, Tokuno2022MNRAS}.
These modes can be classified into two types: inertial modes modified by buoyancy and gravity modes modified by rotation \citep[e.g.][]{Dintrans2000A&A}.
In the Sun's radiative interior, the Brunt-V{\"a}is{\"a}l{\"a} frequency is about three orders of magnitude higher than the rotation rate. Thus, most attention has so far been focused on perturbations of gravity modes by rotation \citep[e.g.][]{Alvan2014A&A}, rather than on inertial modes perturbed by gravity or buoyancy.

In this study, we conducted a linear eigenmode analysis for the solar case, including not only quasi-toroidal inertial modes but also non-toroidal inertial modes (HFR and prograde columnar modes), and examined how they couple to the radiative interior.
Given the high precision and accuracy of solar inertial mode observations, accurate computations of both frequencies and eigenfunctions are essential for interpreting these observations. Coupling to the RZ may potentially influence the eigenmodes restricted to the CZ shell alone. We also studied Rossby modes in the radiative interior, which are perturbed by buoyancy due to the sub-adiabatic stratification.
The paper is organized as follows. In Sect.~\ref{Model_descriptions}, we describe our model of small-amplitude inertial modes, which covers both the CZ and the upper RZ. The eigenfunctions and spectra of the selected inertial modes computed in the setup are analysed in Sect.~\ref{sec:3}. Finally, we discuss the implications of our results in Sect.~\ref{summary}.

\begin{figure*}
    \centering
    \includegraphics[width=0.8\linewidth]{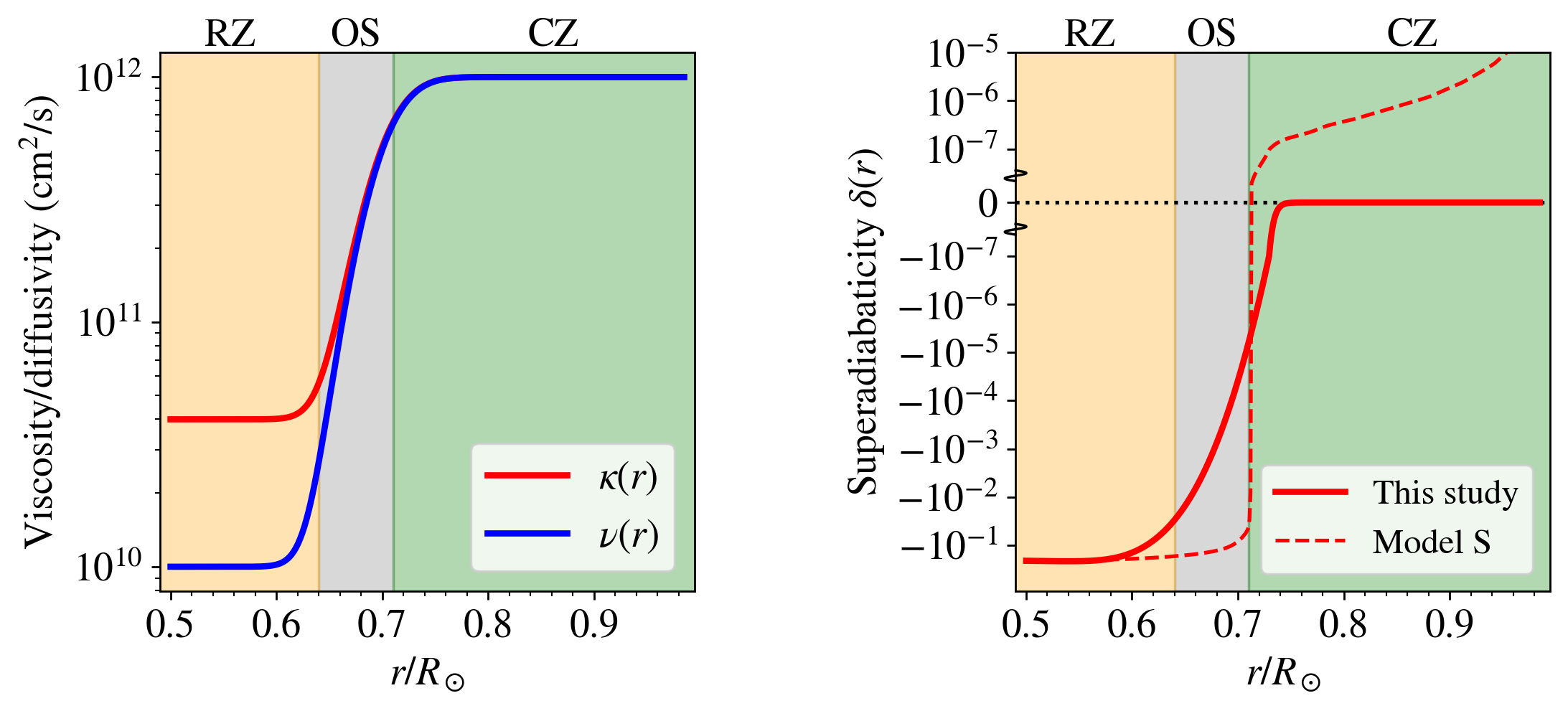}
    \caption{\textit{Left}: Profiles of turbulent viscosity, $\nu(r),$ and turbulent thermal diffusivity, $\kappa(r),$ as functions of radius. They are expressed by Eqs.~\eqref{eq:viscosity} and \eqref{eq:diffusivity}.
    \textit{Right}: Profile of super-adiabaticity, $\delta(r),$ as a function of radius (Eq.~\ref{eq:delta}). 
    The super-adiabaticity profile from the standard Model~S is represented by the dashed red curve \citep{Christensen-Dalsgaard1996Sci}. The $y$-axis is linear for $|\delta|<10^{-7}$ and logarithmic beyond that. 
    In both panels, the green-shaded area denotes the CZ, the grey-shaded region the OS, and the orange-shaded region the RZ.
    }
    \label{fig:2}
\end{figure*}

\section{Inertial modes in a model that combines the CZ and RZ}\label{Model_descriptions}

We used the flexible spectral code \texttt{Dedalus} \citep{Burns2020PRR} to solve the linear eigenvalue problem for low-frequency modes of oscillation in the Sun.
We employed the same approach as in \citet{Mukhopadhyay2025A&A}. 
The setup used in this paper encompasses the upper RZ and CZ of the Sun, with free-surface boundary conditions applied at the top and bottom of the domain. 
We used a fully compressible model, as gravity modes in the radiative interior have been found to be affected by the anelastic approximation \citep[e.g.][]{Brown2012ApJ, Vasil2013ApJ, Hindman2024ApJ}.  

\subsection{Equations for linear oscillations}
We solved the linearized compressible fluid dynamical equations,
\begin{multline}\label{eq:1}
    \frac{\partial \vec{u}}{\partial t}+\vec{v}_0\cdot\vec{\nabla} \vec{u}+\vec{u}\cdot\vec{\nabla} \vec{v}_0+\\
    +2 \Omega_0\,\vec{e}_z\times \vec{u} + \frac{\vec{\nabla}{p_1}}{\rho_0} +\frac{\rho_1}{\rho_0}g\,\vec{e}_r  -  \frac{1}{\rho_0} \vec{\nabla}\cdot \left( \nu \rho_0 \tens{S}\right)=0,
\end{multline}
\begin{equation}\label{eq:2}
     \frac{\partial s_1}{\partial t} +  \vec{v}_0\cdot\vec{\nabla} s_1 +   u_r\frac{\partial s_0}{\partial r} +   \frac{u_{\theta}}{r} \frac{\partial s_0}{\partial \theta}  -\frac{1}{ \rho_0  T_0}\vec{\nabla} \cdot ( \kappa \rho_0  T_0 \vec{\nabla} s_1) =0,
\end{equation}
\begin{equation}\label{eq:3}
    \frac{\partial \rho_1}{\partial t}+\vec{\nabla}\cdot(\rho_0 \vec{u})+\vec{\nabla}\cdot(\rho_1 \vec{v}_0)=0, 
\end{equation}
in combination with the linearized quasi-adiabatic equation of state \citep[e.g.][]{Unno1979book},
\begin{equation}\label{eq:4}
    \frac{p_1}{p_0}=\Gamma_1\frac{\rho_1}{\rho_0}+ \Gamma_1 \nabla_{\rm ad}\frac{\rho_0T_0}{p_0}{s_1} .
\end{equation}
In the above equations, $\vec{u}$, $\rho_1$, $p_1$, and $s_1$ denote the perturbations of velocity, density, pressure, and entropy with respect to the background. We took the background density ($\rho_0$), pressure ($p_0$), temperature ($T_0$), gravitational acceleration ($g$), first adiabatic exponent ($\Gamma_1$), and adiabatic gradient ($\nabla_{\rm ad}$) from the standard solar model, Model~S \citep{Christensen-Dalsgaard1996Sci}. 
The equations were formulated in the Carrington frame of reference rotating at $\Omega_0/2\pi = 456\ \rm nHz$. In this frame, the base flow $\vec{v}_0=(\Omega(r,\theta)-\Omega_0)\, r \sin \theta \, \vec{e}_{\phi}$ denotes the Sun's differential rotation.
The background latitudinal entropy gradient ($\partial s_0/\partial \theta$) is estimated under the assumption of thermal wind balance for solar differential rotation \citep[e.g.][]{Pedlosky1982book, Thompson2003ARA&A}. 
The remaining quantities in the model equations are described in the following subsections.

\subsection{Differential rotation model}

In most of our calculations, we used a simplified analytic profile for the differential rotation given by
\begin{equation}\label{eq:simpleDR}
    \Omega(r, \theta) = \Omega_{\rm CZ}(\theta)+\frac{(\Omega_{\rm RZ} - \Omega_{\rm CZ}(\theta))}{2}\left[1+  { \rm erf}\left(\frac{r_{\rm tc}-r}{d_{\rm tc}} \right)\right] ,
\end{equation}
where $r_{\rm tc}= 0.71\, R_\odot$ and $d_{\rm tc}= 0.045\, R_\odot$ are the position and thickness of the tachocline, respectively.
We assumed that the RZ rotates rigidly at $\Omega_{\rm RZ}/2\pi = 433.5\, \rm nHz$, while the CZ has a rotational shear in latitude as
\begin{equation}
    \Omega_{\rm CZ}(\theta) = \Omega_1 + \Omega_2 \cos^2 \theta + \Omega_3 \cos^4 \theta,
\end{equation}
with $\Omega_1/2\pi=464\, \rm nHz$, $\Omega_2/2\pi=-76\, \rm nHz$, and $\Omega_3/2\pi=-50\, \rm nHz$.
This simplified profile allowed the radial and latitudinal dependences to be separated and reduced memory usage in the \texttt{Dedalus} solver, which enabled us to use a higher grid resolution to better resolve modes with many radial nodes.

However, when modelling the high-latitude modes, we used the actual data of the Sun's differential rotation measured by helioseismology \citep[from][]{Larson2018SoPh}.
This is because the baroclinically unstable high-latitude modes are known to sensitively depend on the differential rotation profile and the associated latitudinal entropy gradient \citep[][]{Gizon2021A&A,Bekki2022A&A2}.
The observed and simplified profiles of the differential rotation are compared in Fig.~\ref{fig:1}. The shaded regions in Fig.~\ref{fig:1}c are described in the following subsection.

\subsection{Turbulent diffusivities and super-adiabaticity}
The strain-rate tensor is given by
\begin{equation}
    \tens{S} =  \vec{\nabla} \vec{u} + \vec{\nabla} \vec{u}^T - \frac{2}{3}(\vec{\nabla} \cdot \vec{u})\ \tens{I}_3 .
\end{equation}
We used radially varying profiles of turbulent viscosity $\nu$ and turbulent thermal diffusivity $\kappa$, similar to \cite{Nandy2002Sci}, given by
\begin{equation}\label{eq:viscosity}
    \nu(r) = \nu_{\rm  RZ} + \frac{(\nu_{\rm CZ}-\nu_{\rm  RZ})}{2}\left[1+  { \rm erf}\left(\frac{r-r_{\rm OS}}{d_{\rm OS}} \right)\right],
\end{equation}
\begin{equation}\label{eq:diffusivity}
    \kappa(r) = \kappa_{\rm  RZ} + \frac{(\kappa_{\rm CZ}-\kappa_{\rm  RZ})}{2}\left[1+  { \rm erf}\left(\frac{r-r_{\rm OS}}{d_{\rm OS}} \right)\right].
\end{equation}
Here, the turbulent viscosity and thermal diffusivity in the CZ ($\nu_{\rm CZ}$, $\kappa_{\rm CZ}$) were both set to be $10^{12}\ \rm cm^2~s^{-1}$. In the RZ below, we adopted the values $\nu_{\rm  RZ}=10^{10}\ \rm cm^2~s^{-1}$ and $\kappa_{\rm RZ}=4\times 10^{10}\ \rm cm^2~s^{-1}$. We assumed a slightly higher thermal diffusivity than turbulent viscosity in the RZ to account for radiative effects \citep[e.g.][]{Brun2011ApJ, Brown2012ApJ}. We used the parameters for the transition depth $r_{\rm OS} = 0.7\, R_\odot$ and the transition thickness $d_{\rm OS}=0.04\, R_\odot$.
The left panel of Fig.~\ref{fig:2} shows the radial profiles of $\nu$ and $\kappa$ used in this study. 

The background radial entropy gradient, $\partial s_0/\partial r$, denotes the degree of convective instability in the background stratification.
Although the RZ is strongly sub-adiabatic (convectively stable), uncertainty remains regarding super-adiabaticity in the bulk CZ.
While Model~S predicts a super-adiabatic stratification throughout the whole CZ, this prediction has been questioned by recent observations and numerical models of solar large-scale convection \citep[e.g.][]{Kapyla2017ApJ,Gizon2021A&A,Hotta2022ApJ,Bekki2024A&A}.
In this study, we employed the following step function for $\partial s_0/\partial r$, which smoothly connects the strongly sub-adiabatic RZ and the adiabatic bulk CZ,
\begin{equation}\label{eq:delta}
     \frac{\partial s_0}{\partial r} = \frac{1}{4} \left( \frac{\partial s_0}{\partial r} \right)_{\rm RZ} \left[ 1- {\rm erf} \left( \frac{r-r_{\delta 1}}{d_{\delta}} \right)\right] \left[ 1 - {\rm erf} \left( \frac{r - r_{ \delta2}}{d_{\delta}}\right) \right] ,
\end{equation}
where $r_{\delta 1} = 0.61 R_\odot$, $r_{\delta 2} = 0.68 R_\odot$, $d_{\delta} = 0.04 R_\odot$, and $(\partial s_0/ \partial r)_{\rm RZ}= 1.32 \times 10^{-2} \ \rm erg \ g^{-1} K^{-1} cm^{-1}$. 
The right panel of Fig.~\ref{fig:2} shows the corresponding radial profile of super-adiabaticity $\delta = - H_p c_p^{-1} (\partial s_0/\partial r )$ used in our computations. 
We note that our $\delta(r)$ profile has a smoother transition between the sub-adiabatic RZ and the adiabatic CZ than that of Model~S.
This is required in our model to avoid oscillations in the projection of the profile onto the Chebyshev basis.

Based on the profiles of turbulent diffusivities and super-adiabaticity shown in Fig.~\ref{fig:2}, our numerical domain can be divided into three regions:
\begin{itemize}
    \item the RZ ($r < 0.64R_{\odot}$), where turbulent motions are suppressed ($\nu$ and $\kappa$ are substantially small), and the stratification is strongly sub-adiabatic ($\delta \sim -0.1$),
    \item the overshooting (OS) layer ($0.64R_{\odot} \leq r \leq 0.71R_{\odot}$), where turbulent motions penetrate but the stratification is significantly sub-adiabatic \citep[e.g.][]{Hotta2017ApJ, Kapyla2017ApJ},
    \item the CZ ($r > 0.71R_{\odot}$) where the vigorous convective motions lead to significant values of $\nu$ and $\kappa$ and the stratification is close to adiabatic ($|\delta| \lesssim 10^{-6}$).
\end{itemize}
These three regions are highlighted by different colours in Fig.~\ref{fig:2}. Figure~\ref{fig:1}c demonstrates that the tachocline with significant radial differential rotation overlaps with the OS defined in our setup.

\subsection{Free-surface boundary conditions}\label{sec:BC}
We set the lower boundary in the upper RZ at $r_{\rm i} = 0.5\, R_\odot$ and the upper boundary slightly below the photosphere at $r_{\rm o} = 0.985\, R_\odot$.
At both radial boundaries, in contrast to the conventionally used impenetrable boundary condition, we imposed the free-surface boundary condition for the radial velocity $u_r$. This is done in the presence of base flow $\vec{v}_0$ by setting the Lagrangian derivative of total pressure to $0$ \citep[e.g.][]{Mei2005Book}, as
\begin{equation}
    \frac{D }{D t} (p_0+p_1) \approx \frac{\partial  p_1}{\partial t}+ \vec{u}\cdot \vec{\nabla} p_0 + \vec{v}_0\cdot \vec{\nabla} p_1 =0.
\end{equation}
Using the wave Ansatz where all the perturbations are assumed to be proportional to $\exp(\ii m\phi - \ii\omega t)$, this translates to
\begin{equation}\label{Lagrangian_pressure}
    -\ii\omega p_1-u_r\rho_0 g + \ii m (\Omega(r,\theta) - \Omega_0)  p_1 =0. 
\end{equation}
Equation~\eqref{Lagrangian_pressure} was forced at both radial boundaries.
We validated this free-surface boundary condition by reproducing the dispersion relation of the $f$ modes on a spherical surface \citep[e.g.][]{Christensen-Dalsgaard2002RvMP, Antia1998A&A}.
For low-frequency inertial modes whose oscillation periods are much longer than those of $f$ modes, the free-surface boundary condition is found to almost reduce to the impenetrability condition \citep[e.g.][]{Hindman2022ApJ, Hindman2023ApJ}. 
Nevertheless, the free-surface boundary condition is expected to work better for $g$~modes and gravito-inertial modes in the RZ. 
In addition, we applied the conventional horizontal stress-free condition, i.e. $\tens{S}_{r\theta}=\tens{S}_{r\phi}=0$, at both radial boundaries. We assumed that the flux of entropy perturbations vanishes at the bottom boundary, and entropy perturbations vanish at the top boundary. 
We note that this treatment is consistent with the free-surface boundary condition, given that the radial entropy gradient $\partial s_0/\partial r$ vanishes at the top boundary in our setup.

\subsection{Eigenvalue solver}
We followed the same approach as in \citet{Mukhopadhyay2025A&A} to solve the eigenvalue problem of the governing equations (Eqs.~\ref{eq:1}--\ref{eq:4}) with the boundary conditions (Sect.~\ref{sec:BC}),
considering the wave Ansatz 
$\exp(\ii m\phi - \ii \omega t)$
for each azimuthal order $m$. We solved the sparse problem on the spherical shell basis of \texttt{Dedalus} and selected the modes of interest, in the same way as in \citet{Mukhopadhyay2025A&A}. 
We used a typical resolution of 180 points in radius and 60 points in latitude in calculations with the simplified differential rotation. When the observed solar differential rotation was adopted, we used a grid with 90 radial points and 36 latitudinal points. 
All eigenmodes presented in the paper are numerically converged to errors well below the observational uncertainties \citep{Gizon2021A&A}.

\begin{figure*}
    \centering
    \includegraphics[width=0.86\linewidth]{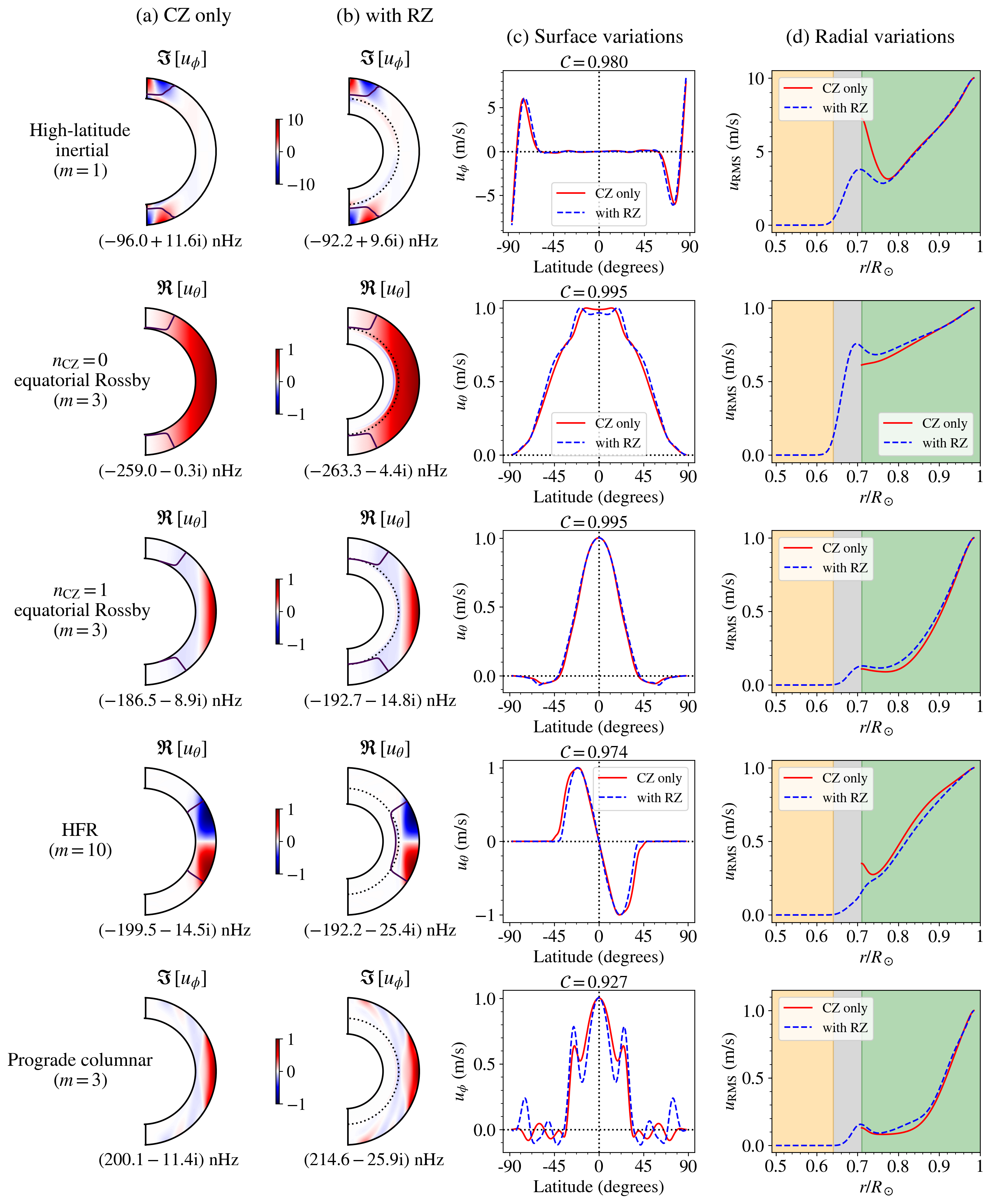}
    \caption{Comparison of selected inertial modes computed in the setups that include or exclude the RZ. (a) Velocity eigenfunctions in meridional cross-sections from the CZ-only model. 
    The real part of the eigenfunction corresponds to a longitude $\phi_0$ (where $u_\theta$ is maximum), while the imaginary part corresponds to the longitude $\phi_0-\pi/2m$.
    The $m=1$ high-latitude mode is normalized to have the maximum surface velocity of $10\, \rm m~s^{-1}$, while the other modes are normalized to have a maximum surface velocity of $1\, \rm  m~s^{-1}$.
    Solid black curves denote the critical layers of the modes, where $\Re [\omega] = m (\Omega-\Omega_0)$.
    The frequencies are measured in the Carrington frame.
    (b) Same as panel a, but from the extended model that includes the RZ. Dotted black lines denote the base of the CZ.
    (c) Horizontal velocity eigenfunctions as functions of latitude at the surface. Dashed blue and solid red curves represent the results with and without the RZ, respectively. The correlation coefficient between the eigenfunctions, $\mathcal{C}$, defined in Eq.~\eqref{eq.corr}, is noted above each subplot.
    (d) Radial profiles of the RMS velocity. 
    }
    \label{fig:3}
\end{figure*}

\begin{figure}
    \centering
    \includegraphics[width=0.95\linewidth]{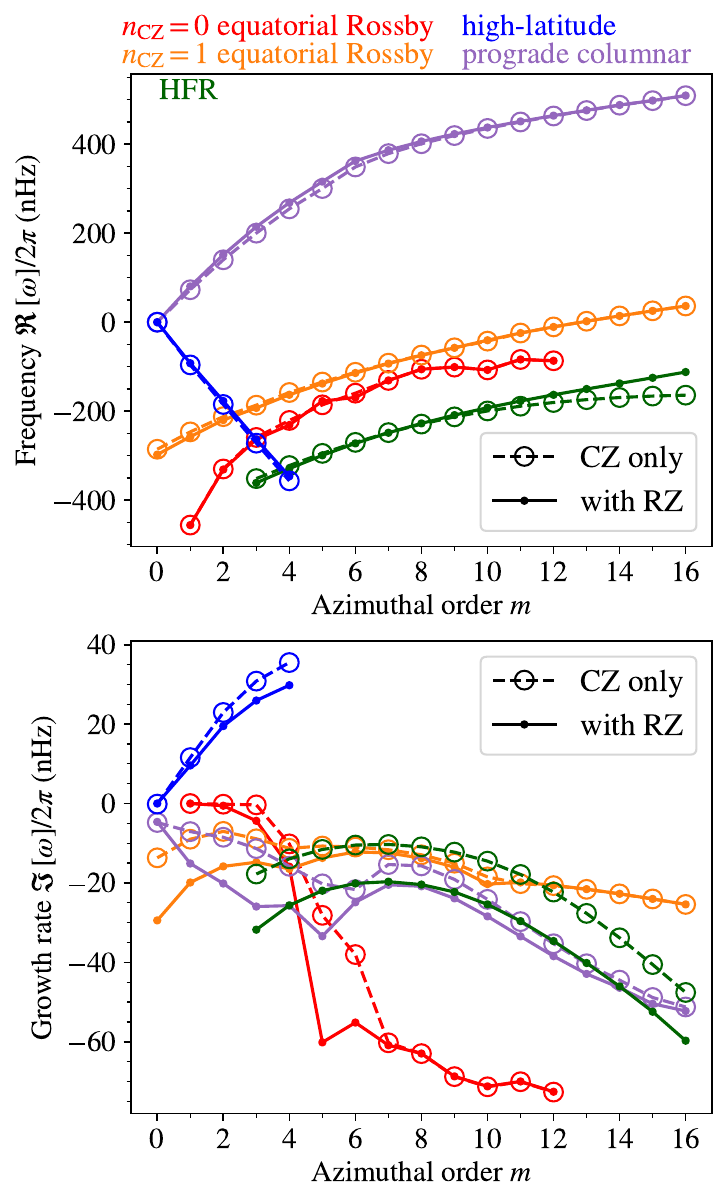}
    \caption{\textit{Top}: Dispersion relations of the inertial modes in the CZ computed in the setups that include the RZ (solid lines with points) and exclude the RZ (dashed lines with open circles). The colours denote the various types of inertial modes. 
    \textit{Bottom}: Growth rates of the same modes, with the same notations.
    }
    \label{fig:4}
\end{figure}

\begin{figure}
    \centering
    \includegraphics[width=0.95\linewidth]{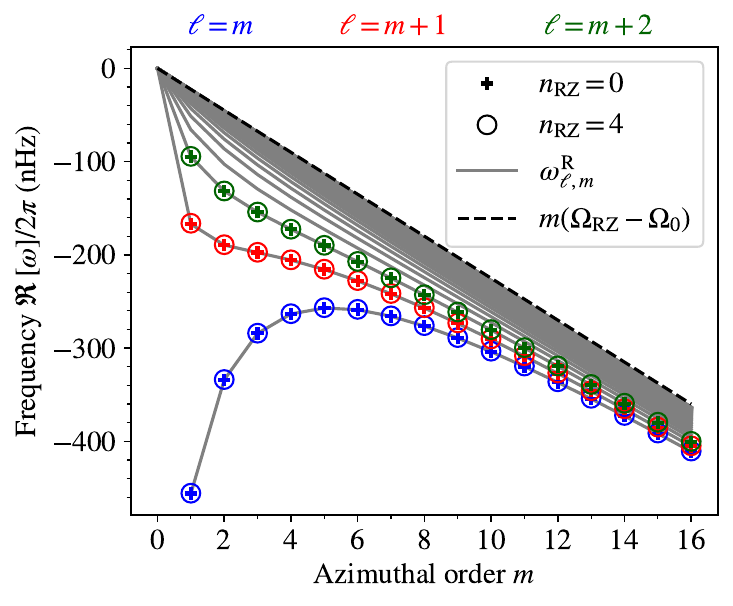}
    \caption{Dispersion relations of the Rossby modes inside the RZ with $\ell=m$ (navy blue),  $\ell = m+1$ (red), and $\ell =m+2$ (dark green) for azimuthal orders $1\leq m \leq 16$. The frequencies are measured in the Carrington frame. Plus symbols ($n_{\rm RZ}=0$) and open circles ($n_{\rm RZ}=4$) denote modes with different numbers of radial nodes, $n_{\rm RZ}$, in the region $0.5R_\odot\leq r\leq 0.71 R_\odot$. Solid grey curves represent the theoretical dispersion relations of the classical Rossby modes $\omega^{\rm R}_{\ell,\, m}$ given by Eq.~\eqref{eq:Rossby_dispersion_RZ}. The dashed black line denotes the maximum possible value of $\omega^{\rm R}_{\ell,\, m}$.}
    \label{fig:5}
\end{figure}

\begin{figure*}
    \centering
    \includegraphics[width=0.9\linewidth]{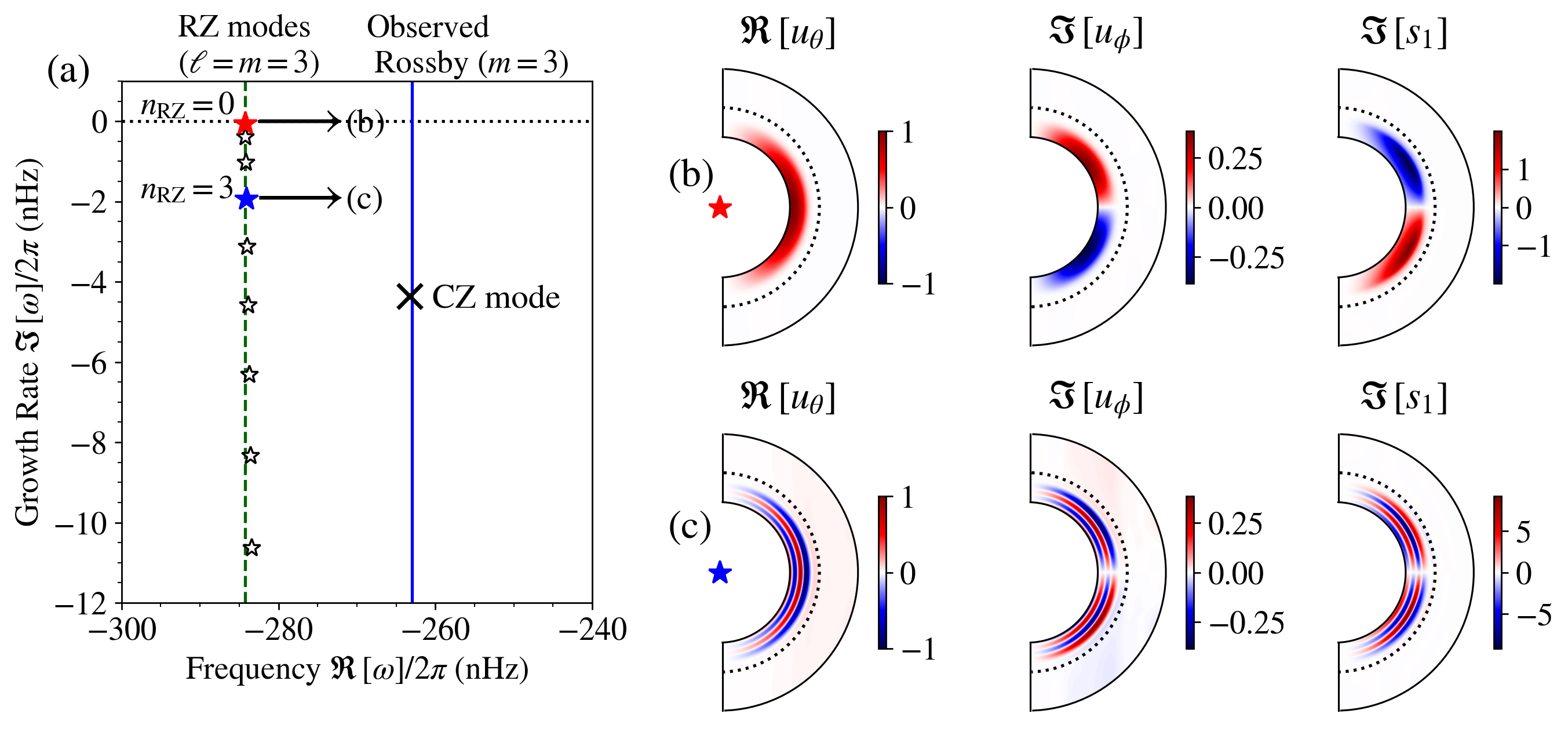}
    \caption{Position of the RZ Rossby modes with $\ell=m=3$ in the complex frequency space and their eigenfunctions.
    (a) Complex frequency spectrum of equatorial Rossby modes in the RZ with azimuthal order $m=3$. Star symbols denote the RZ Rossby modes, while the cross symbol denotes the CZ Rossby mode. 
    The red (blue) star marks the RZ Rossby mode with the number of radial nodes, $n_{\rm RZ}=0$ ($n_{\rm RZ}=3$), in the region $0.5R_\odot\leq r\leq 0.71 R_\odot$. 
    The vertical solid blue line denotes the observed Rossby mode frequency, while the vertical dashed green line denotes the frequency given by Eq.~\eqref{eq:Rossby_dispersion_RZ}. 
    (b) Meridional eigenfunction of the RZ Rossby mode with $n_{\rm RZ}=0$ (red star). The longitudes corresponding to the real and imaginary components are designated in the same way as in Fig.~\ref{fig:3}. Dotted lines denote the position of the base of the CZ. Eigenfunctions are normalized such that the maximum of $u_\theta$ is $1\,\rm m~s^{-1}$. The unit of $s_1$ is $\rm erg\, g^{-1}\, K^{-1}$.
    (c) Same as panel b but for the RZ Rossby mode with $n_{\rm RZ}=3$ (blue star). }
    \label{fig:6}
\end{figure*}

\begin{figure*}
    \centering
    \includegraphics[width=0.9\linewidth]{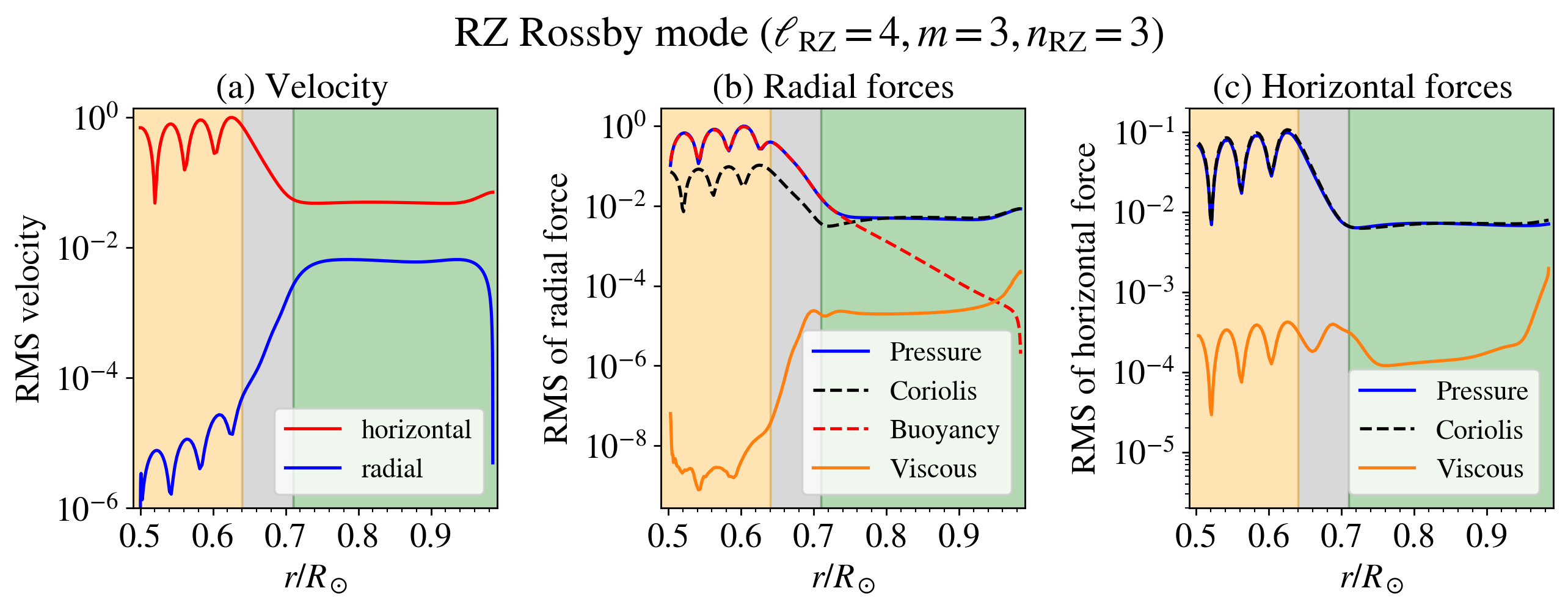}
    \caption{Analyses of the non-sectoral Rossby mode in the RZ with $\ell_{\rm RZ}=4, m=3, n_{\rm RZ}=3$. (a) Radial profiles of the RMS velocity in the horizontal (red) and radial (blue) directions. The profiles are normalized to have a maximum total RMS velocity of unity.
    (b)~RMS of the radial component of various forces, defined in Eqs.~\eqref{eq:Forces1}~--~\eqref{eq:Forces4}. The pressure gradient force (solid blue), Coriolis force (dashed black), buoyancy force (dashed red), and viscous force (solid orange) are normalized so that the maximum of the pressure gradient force is unity. (c)~Same as panel b, but for the horizontal component of the forces. Colours of the shaded regions follow the same notation as in Fig.~\ref{fig:2}.}
    \label{fig:7}
\end{figure*}

\begin{figure*}
    \centering
    \includegraphics[width=0.95\linewidth]{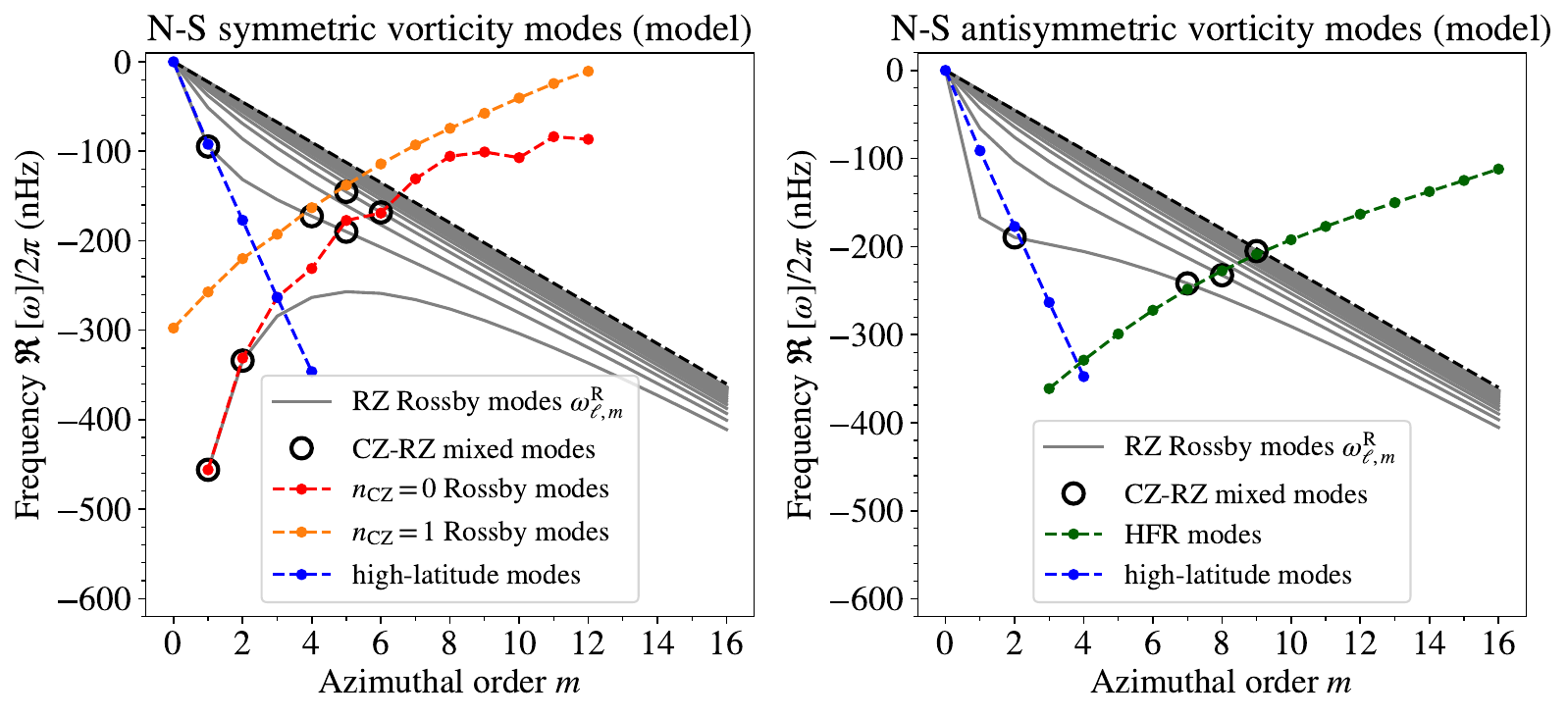}
    \caption{Frequencies of the retrograde-propagating inertial modes in the CZ (dashed coloured; the same as in Fig.~\ref{fig:4}) overplotted with the dispersion relations of the RZ Rossby modes (solid grey; given by Eq.~\eqref{eq:Rossby_dispersion_RZ}).
    The left and right panels correspond to the modes with north--south symmetric and antisymmetric vorticity, respectively.
    Open circles mark the subset of RZ Rossby modes that can form mixed modes with the CZ inertial modes.
    }
    \label{fig:8}
\end{figure*}

\begin{figure}
    \centering
    \includegraphics[width=0.95\linewidth]{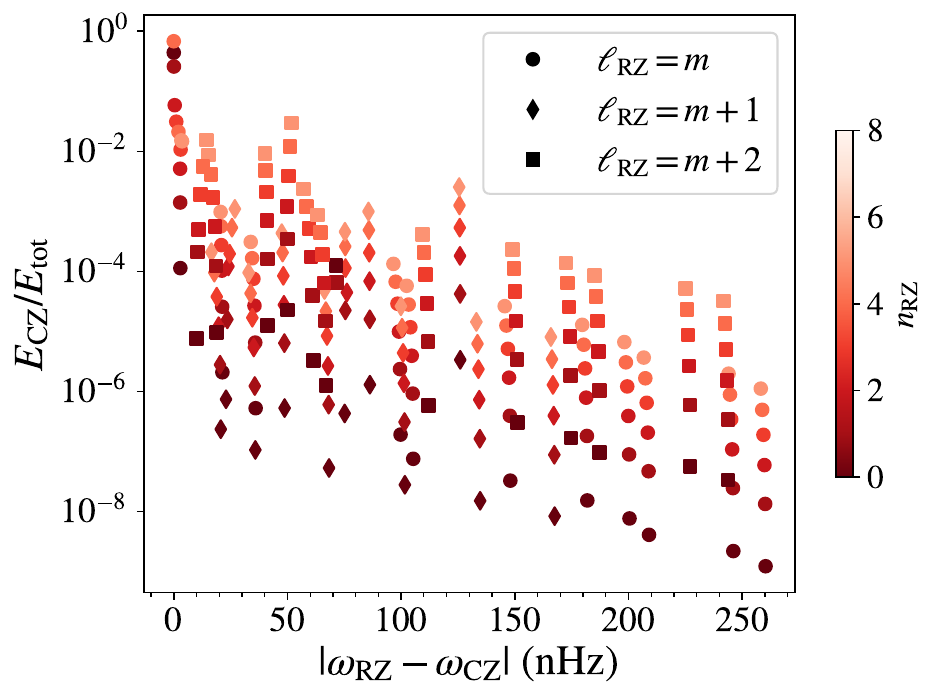}
    \caption{Fraction of kinetic energy density, $E=\langle \rho_0 u^2\rangle$, of the RZ Rossby modes contained in the CZ as a function of the absolute difference between the eigenfrequency of each mode and that of the nearest CZ inertial mode (see Fig.~\ref{fig:8}). For clarity, we show only the RZ Rossby modes with $\ell_{\rm RZ}=m$ (circles), $\ell_{\rm RZ}=m+1$ (diamonds), and $\ell_{\rm RZ}=m+2$ (squares). The colour of the symbols denotes the number of radial nodes ($n_{\rm RZ}$) in the region $0.5R_\odot\leq r\leq 0.71 R_\odot$. Only modes with $n_{\rm RZ}\leq 5$ are shown. 
    }
    \label{fig:9}
\end{figure}

\begin{figure*}
    \centering
    \includegraphics[width=0.917\linewidth]{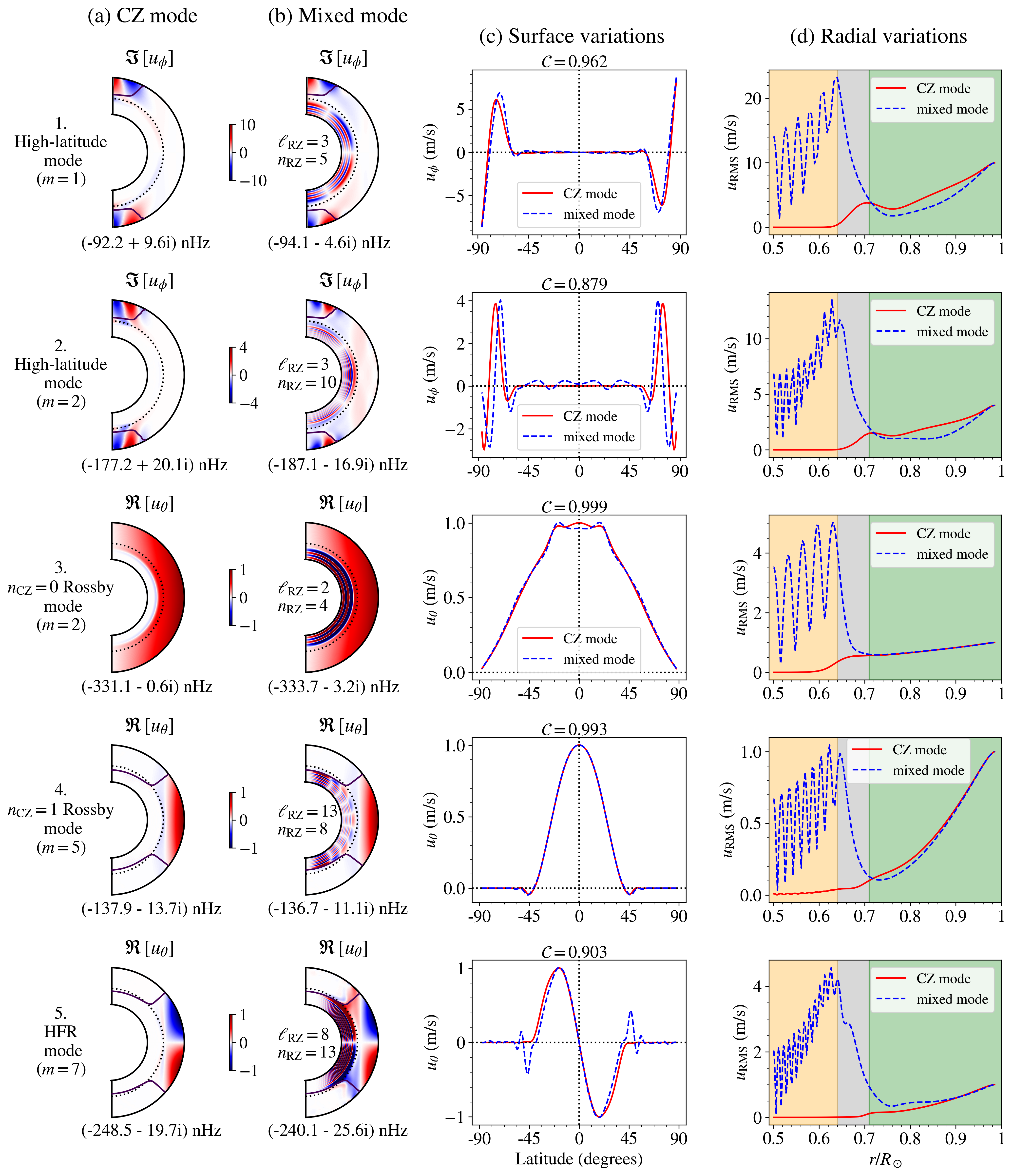}
    \caption{Selected inertial modes computed in the extended setup, which includes the RZ. (a) Modes with motions predominantly in the CZ and decaying in the RZ. (b) Modes with comparable motions in the RZ and CZ, with flows in the CZ similar to those of the corresponding modes in panel (a). (c) and (d): Surface and radial variations of the modes, respectively, as in Fig.~\ref{fig:3}. 
  The spherical harmonic degree and the number of radial nodes of the RZ Rossby modes in the region $0.5R_\odot\leq r\leq 0.71 R_\odot$ are denoted by $\ell_{\rm RZ}$ and $n_{\rm RZ}$, respectively. The $m=1$ high-latitude modes are normalized to have a maximum surface velocity of $10~\rm m~s^{-1}$, the $m=2$ high-latitude modes to $4~\rm m~s^{-1}$, and the other modes to $1~\rm m~s^{-1}$. The notation and layout are the same as in Fig.~\ref{fig:3}.
    }
    \label{fig:10}
\end{figure*}

\section{Inertial modes of the extended model}\label{sec:3}
\subsection{Modes confined primarily to the CZ}\label{sec:3.1}
First, we examined how the inertial modes of the CZ are affected by the presence of the RZ below.
To this end, we performed calculations with and without the RZ and compared the resulting eigenmodes.
The model without the RZ has the lower boundary at $r_{\rm i}=0.71\, R_\odot$ with an impenetrable boundary condition, thereby excluding both the OS and the RZ. This CZ-only model has been traditionally employed in previous studies \citep{Bekki2022A&A, Mukhopadhyay2025A&A}. 
Except for this lower boundary treatment, the two models are identical.

Figures~\ref{fig:3}a and \ref{fig:3}b compare the eigenfunctions of the $m=1$ high-latitude mode, $m=3$ equatorial Rossby modes with no radial node ($n_{\rm CZ}=0$) and with one radial node in the CZ ($n_{\rm CZ}=1$), the $m=10$ HFR mode, and the $m=3$ prograde columnar mode.
All of these modes have already been discussed in \cite{Mukhopadhyay2025A&A} within the CZ-only framework.
We find that the inclusion of the RZ has a limited impact on the eigenfunctions of these inertial modes within the CZ.
In fact, these modes are found to decay in the RZ (Fig.~\ref{fig:3}b).
Figure~\ref{fig:3}c compares the surface velocity eigenfunctions between the two models, where the eigenfunctions are normalized such that the modes have the same total horizontal velocity power at the surface in the two models.
The differences in the velocity eigenfunctions at the surface ($r=r_{\rm o}$) can be measured by the correlation coefficient 
\begin{equation}
    \mathcal{C} = \frac{\int_S (u_{\theta,\, \rm RZ\,} u_{\theta,\, \rm CZ\,} + u_{\phi,\, \rm RZ\,} u_{\phi,\,\rm CZ\,})\, dS}{ \sqrt{\int_S (u_{\theta,\,\rm RZ\,}^2 + u_{\phi,\,\rm RZ\,}^2) \, dS} \sqrt{\int_S (u_{\theta,\,\rm CZ\,}^2 + u_{\phi,\,\rm CZ\,}^2)\, dS}},
    \label{eq.corr}
\end{equation}
where ${S}$ represents the spherical surface and $\vec{u}_{\rm RZ}$ and $\vec{u}_{ \rm CZ}$ are from the models with and without the RZ, respectively.
We find that {$\mathcal{C} \gtrsim 0.9$} for the inertial modes discussed here, demonstrating that the way these inertial modes are observed at the top of the CZ is almost unaffected by the inclusion of the RZ.
Figure~\ref{fig:3}d compares the radial variations of mode power measured by root-mean-square (RMS) velocity $u_{\rm RMS}$.
It is shown that the power distribution in the bulk CZ is only marginally affected by the inclusion of the RZ, except near the base of the CZ, where the deviation can become significant (e.g. $m=1$ high-latitude mode). Although the modes decay in the RZ, they can exhibit finite motion in the OS. This can affect mode frequencies and growth rates by altering mode dissipation (see Appendix~\ref{Appendix:dissipation}).
Figure~\ref{fig:3} suggests that CZ inertial modes are generally decoupled from the RZ, as modes with finite radial motions are excluded from the strongly sub-adiabatic RZ. Although purely toroidal modes can form global eigenfunctions under uniform rotation (Appendix~\ref{Appendix:Uniform-rot}), the inclusion of differential rotation inevitably makes the CZ modes decoupled from the RZ. Thus, we note that the reported decoupling of CZ inertial modes from the RZ is a generic result, largely independent of model details such as viscosity profiles and lower boundary conditions.

Figure~\ref{fig:4} compares the frequencies and growth rates of the above-discussed inertial modes in the two setups that include and exclude the RZ. 
The inclusion of RZ affects the mode frequencies by at most a few nanohertz ($< 10 \rm ~nHz$) for most inertial modes except some non-toroidal modes (HFR modes at $m\geq 12$ and prograde columnar modes at $m\leq 6$). The differences in frequencies are remarkably large ($>20~\rm nHz$) only for HFR modes at $m>12$.
We note that these HFR modes are special in several aspects: they are strongly non-toroidal, and their radial motions tend to be more localized near the base of the CZ as $m$ increases \citep[][]{Bekki2024A&A}. Furthermore, their critical layers penetrate into the OS (see Fig.~\ref{fig:3}b). These various factors change the effective rotation rate associated with the mode and cause an apparent shift in the frequency measured in the Carrington frame (refer to Appendix~\ref{Appendix:HFR} for details). This Doppler shift of the frequency in the Carrington frame is further amplified for high $m$.
In contrast to \citet{Hindman2023ApJ}, the dispersion relation of the prograde columnar modes in our setup is not significantly affected by the inclusion of the RZ because the strongly sub-adiabatic RZ itself acts like a lower boundary to these convective modes. 

Although the changes in frequencies are rather insignificant for most inertial modes, the inclusion of the RZ has a stronger impact on the mode growth rates.
As shown in the bottom panel of Fig.~\ref{fig:4}, the modes tend to be more damped when the RZ is included.
This results from the dissipation in the OS caused by the inclusion of the RZ (see Fig.~\ref{fig:A1}). For high $m$, the modes tend to be confined to the outer part of the CZ, so their growth rates are not significantly affected by the inclusion of the RZ. This is not the case for the HFR modes, which still exhibit significant motions throughout the CZ at $m=16$, as their critical layers are located near the CZ base.

\subsection{Modes confined primarily to the RZ} \label{sec:3.2}
Next, we discuss the modes that have dominant motions in the RZ.
We find that there exist quasi-toroidal inertial modes trapped inside the RZ whose eigenfunctions are well represented by the spherical harmonics and follow the dispersion relations of the classical Rossby modes
\begin{equation}\label{eq:Rossby_dispersion_RZ}
    \omega^{\rm R}_{\ell,\, m}=-\frac{2m\Omega_{\rm RZ}}{\ell (\ell+1)} + m(\Omega_{\rm RZ} - \Omega_0),
\end{equation}
where $\ell$ is the harmonic degree. Here, the second term represents the Doppler frequency shift in the Carrington frame.
For each combination of ($\ell,m$), these RZ Rossby modes can have any number of radial nodes $n_{\rm RZ}$ while sharing the same frequencies $\omega^{\rm R}_{\ell,\, m}$, where $n_{\rm RZ}$ is counted for $u_\theta$ in the region $0.5R_\odot\leq r\leq 0.71 R_\odot$.
Figure~\ref{fig:5} displays their dispersion relations for $\ell-m=0,\ 1,$ and $2$, with two different $n_{\rm RZ}$, clearly demonstrating that the computed frequencies agree strikingly well with the theoretical prediction (Eq.~\ref{eq:Rossby_dispersion_RZ}) and that the frequencies are independent of $n_{\rm RZ}$.

Figure~\ref{fig:6}a shows the complex eigenfrequency spectrum at $\ell=m=3$.
The equatorial Rossby mode inside the CZ, discussed in Sect.~\ref{sec:3.1}, can be identified at $\Re[\omega]/2\pi=-263.3$~nHz (denoted by the cross symbol). 
Apart from that, near the classical Rossby mode frequency $\omega^{\rm R}_{\ell=m=3}$, there exist many RZ Rossby modes with different $n_{\rm RZ}$ (denoted by star symbols).
Their growth rates decrease with the increase in $n_{\rm RZ}$ as dissipation increases with $n_{\rm RZ}$ (see Appendix~\ref{Appendix:dissipation}).
Figures~\ref{fig:6}b and \ref{fig:6}c further show the meridional eigenfunctions of the RZ Rossby modes with $n_{\rm RZ}=0$ and $3$, respectively. 
The eigenfunctions indicate that the modes are strongly confined within the RZ and decay in the CZ.
Since the RZ is uniformly rotating and no critical layers exist, the horizontal eigenfunctions in the RZ show no major deviations from the corresponding spherical harmonics.
We also note that these RZ Rossby modes have substantial entropy perturbations arising from the strongly sub-adiabatic background stratification.

As has been reported in \citet{Blume2024ApJ}, the Rossby modes in the sub-adiabatic RZ can have any allowed combination of ($\ell,m$), including the non-sectoral ($\ell \ne m$) components (see also Fig.~\ref{fig:5}).
This is in striking contrast to the Rossby modes in the adiabatically stratified CZ, where only sectoral ($\ell=m$) modes can exist, as only sectoral modes can sustain the required radial force balance between the Coriolis and pressure gradient forces \citep{Provost1981A&A, Damiani2020A&A}. 
In the sub-adiabatic RZ, on the other hand, the radial component of the Coriolis force can be balanced by the combined effect of buoyancy and pressure to maintain the toroidal nature of these Rossby modes.
To confirm this, we analysed the force balance in the RZ Rossby modes. The various forces are given by
\begin{align}
\label{eq:Forces1}
    & \vec{F}_{\rm Coriolis} = -2\Omega_0 (\vec{e}_z \times \vec{u}),\\
    & \vec{F}_{\rm Pressure} = -\vec{\nabla}\left( \frac{p_1}{\rho_0}\right),\\
    & \vec{F}_{\rm Buoyancy} = \frac{s_1}{c_p}g \vec{e}_r,\\ \label{eq:Forces4}
    & \vec{F}_{\rm Viscous} = \frac{1}{\rho_0} \vec{\nabla}\cdot \left(\nu \rho_0 \tens{S}\right).
\end{align}
Figure~\ref{fig:7} presents the force balance analysis for a non-sectoral RZ Rossby mode with $(\ell, m )=(4,3)$. 
Figure~\ref{fig:7}a shows that the mode is quasi-toroidal, where the radial motions are several orders of magnitude smaller than the horizontal motions. 
However, the small radial motions induce a substantial buoyancy force in the sub-adiabatic RZ to balance the radial pressure gradient and Coriolis forces (Fig.~\ref{fig:7}b). Nevertheless, the geostrophic nature of the traditional Rossby modes persists in the horizontal force balance (Fig.~\ref{fig:7}c). This causes the Rossby modes in the RZ to have frequencies determined entirely by the rotation rate, given by Eq.~\eqref{eq:Rossby_dispersion_RZ}.

\subsection{CZ--RZ mixed modes}\label{sec:3.3}
We now explore modes with significant motion in both the CZ and RZ.
Although we call them CZ--RZ mixed modes in this paper, they correspond to low-frequency modes arising from the mixing between Rossby modes in the RZ and inertial modes in the CZ, and should not be confused with the $p$~--~$g$ mixed modes commonly discussed in asteroseismology.
In practice, we find that significant mode mixing only occurs when $|\Delta \omega|\lesssim 10\, \rm nHz$, where $|\Delta \omega|$ is the frequency separation between the RZ Rossby mode and the nearest CZ inertial mode, consistent with avoided-crossing behaviour.
Such mode-mixing can occur only when the RZ Rossby mode and its nearest CZ inertial mode have the same north--south symmetry.
In such special cases, we encounter CZ--RZ mixed modes, whose eigenfunctions have substantial amplitudes in both the CZ and RZ. 
A caveat should be noted: Although we refer to them as mixed modes, they are still essentially RZ Rossby modes that have non-negligible amplitudes in the CZ, and only a subset of RZ Rossby modes are allowed to behave in this way.
Figure~\ref{fig:8} highlights $11$ examples illustrating how mode mixing between the RZ and CZ can occur in the dispersion diagrams for both equatorially symmetric and antisymmetric modes. 
We note that the possible options of mode mixing are quite limited, as the frequencies of the RZ Rossby modes are restricted to a finite range by Eq.~\eqref{eq:Rossby_dispersion_RZ}.
For instance, the mixing between the $\ell_{\rm RZ}=m$ Rossby modes in the RZ and the $n_{\rm CZ}=0$ equatorial Rossby modes in the CZ can only occur at $m=1$ and $2$. Here, $\ell_{\rm RZ}$ denotes the harmonic degree of the RZ Rossby component of these mixed modes.
The $n_{\rm CZ}=0$ Rossby modes in the CZ with $m=5$ and $6$ can also couple with the RZ Rossby modes but with $\ell_{\rm RZ}-m = 2$ and $6$, respectively. 
A similar coupling is allowed for the $n_{\rm CZ}=1$ Rossby modes in the CZ with $m=4$ and $5$. 
As for the high-latitude modes, only the $m=1$ mode with north--south symmetric vorticity and the $m=2$ mode with the opposite symmetry can couple with the RZ Rossby modes.
Finally, the HFR modes with $m=7,8,$ and $9$ are allowed to couple with the RZ Rossby modes with north--south antisymmetric vorticity.
{The mode masses of all these mixed modes are, however, dominated by the RZ contribution (as quantified in Appendix~\ref{Appendix:mode-mass}).}
We note that there is no mixed mode between the retrograde-propagating RZ Rossby modes and the prograde columnar modes.

In principle, CZ--RZ mixed modes are allowed to have any number of radial nodes $n_{\rm RZ}$ in the region $0.5R_\odot\leq r\leq 0.71 R_\odot$, since the frequencies of the RZ Rossby modes are independent of $n_{\rm RZ}$.
However, $n_{\rm RZ}$ has a significant impact on how the mode power is distributed between the CZ and RZ: 
As $n_{\rm RZ}$ increases, the relative velocity amplitude in the CZ increases. 
A possible explanation for this is given in Appendix~\ref{Appendix:dissipation}.
Figure~\ref{fig:9} displays the normalized kinetic energy within the CZ of the RZ Rossby modes as a function of the absolute frequency difference $|\Delta\omega|$ between each RZ Rossby mode and its nearest CZ inertial mode.
It clearly shows that the mode tends to be more strongly coupled and has more power inside the CZ when $|\Delta\omega|$ is smaller.
It is also demonstrated that the fraction of kinetic energy in the CZ increases with $n_{\rm RZ}$.
These results collectively indicate that, for an RZ Rossby mode to be a CZ--RZ mixed mode, it must have high $n_{\rm RZ}$ as well as a frequency sufficiently close ($|\Delta \omega|\lesssim 10\, \rm nHz$) to that of the corresponding CZ inertial mode.
Note that the threshold $|\Delta\omega|\lesssim 10$ nHz for coupling is a rough estimate, phenomenologically obtained for the setup using Fig.~\ref{fig:9}.

The velocity eigenfunctions of selected CZ--RZ mixed modes are compared in Fig.~\ref{fig:10} with those of the CZ inertial modes (discussed in Sect.~\ref{sec:3.1}), showing overall remarkable similarity in the CZ.
Notably, the surface eigenfunctions exhibit only minor differences, suggesting that it is challenging to determine from surface observations alone whether the mode is mixed with RZ Rossby modes or not.
As a robustness check, we verified numerical convergence for two representative CZ--RZ mixed modes that exhibit strong spatial variations in radial and latitudinal directions (see Appendix~\ref{Appendix:resolution}).

\begin{table}
\caption{Mode mass of the CZ and mixed inertial modes from Fig.~\ref{fig:10}.}
    \centering
    \begin{tabular}{l c c c}
    \hline \hline
        Modes     & Mode mass   &  Frequency & Mode type  \\ from Fig.~\ref{fig:10}
        & ($10^{-3}M_\odot$) & (nHz) & (in CZ)
        \\ \hline 
        1. (a) CZ & $5.9$ & $-92.2$  & 
        \multirow{2}{*}{high-lat. $m=1$}\\
         1. (b) mixed & $163.5$ & $-94.1$  \\ \hline
        2. (a) CZ & $5.5$ & $-177.2$ & \multirow{2}{*}{high-lat. $m=2$} \\
         2. (b) mixed & $317.9$ & $-187.1$ \\ 
         \hline
        3. (a) CZ & $16.2$ & $-331.1$ & \multirow{2}{*}{Rossby $m=2$} \\
         3. (b) mixed & $773.7$ & $-333.7$ &  \\
             \hline
         4. (a)  CZ & $2.7$ & $-137.9$ & \multirow{2}{*}{Rossby $m=5$ } \\
         4. (b) mixed & $34.9$ &  $-136.7$ & \\
             \hline
         5. (a) CZ & $2.6$ & $-248.5$ & \multirow{2}{*}{HFR $m=7$}\\
         
         5. (b) mixed & $594.4$ & $-240.1$ \\
             \hline
    \end{tabular}
    \label{tab:1}
\end{table}

\section{Discussion}\label{summary}
\subsection{Most CZ modes are weakly affected by the RZ}
In this work, we extended the linear model of inertial modes in the Sun's differentially rotating CZ of \citet{Mukhopadhyay2025A&A} to include the sub-adiabatic RZ.
To study how the RZ affects the solar inertial modes in the CZ, we first compared the eigenmodes computed in models that include and exclude the RZ.
We find that the mode frequencies are only marginally affected (a few nanohertz) by the inclusion of the RZ, besides some non-toroidal modes (HFR modes at $m\geq 12$ and prograde columnar modes at $m\leq 6$).
In the model with the RZ, the eigenfunctions are found to slowly decay in the OS before vanishing in the RZ.
The associated viscous dissipation in the OS leads to larger damping rates of these inertial modes. 
Within the CZ, the eigenfunctions are primarily altered near the base of the CZ. 
On the other hand, the surface eigenfunctions are shown to be almost insensitive to the inclusion of the RZ.
Therefore, while the RZ introduces only minor modifications, the CZ-only model remains a valid and practical framework for studying the dispersion relations and surface eigenfunctions of the solar inertial modes and for interpreting the solar observations. 
If very accurate frequencies and eigenfunctions are desired, extending the computational domain down to $0.55 R_\odot$ will suffice as the eigenfunctions vanish by $0.6R_\odot$ (see Fig.~\ref{fig:3}).
Finally, it should be noted that we used a more gradually varying super-adiabaticity profile than that predicted by Model~S, corresponding to a thicker OS layer. A more realistic OS model will be needed for better modelling.

\subsection{RZ Rossby modes and their coupling with CZ modes}
As found by \citet{Blume2024ApJ}, the sub-adiabatic and uniformly rotating RZ can host a wide variety of quasi-toroidal Rossby modes, including the non-sectoral ones (which do not exist in the quasi-adiabatic CZ).
We find that they follow the classical Rossby mode dispersion relation (Eq.~\ref{eq:Rossby_dispersion_RZ}) and have horizontal eigenfunctions well represented by spherical harmonics.
They can have all possible combinations of ($\ell_{\rm RZ},m,n_{\rm RZ}$), because the radial force balance can be maintained with the help of buoyancy. 
However, since their frequencies are entirely dictated by their rotation rate, they should be interpreted as inertial modes perturbed by buoyancy. The perturbations in structure and frequency of the modes due to buoyancy are rather minor because the Brunt-V{\"a}is{\"a}l{\"a} frequency is significantly larger than the rotation rate \citep{Dintrans2000A&A}.

As the number of radial nodes ($n_{\rm RZ}$) increases, the RZ Rossby modes are more damped and, in turn, have more power in the CZ (see Appendix~\ref{Appendix:dissipation}). 
We find that, when the frequencies of the RZ Rossby modes are within $\sim 10\,\rm nHz$ of the nearest CZ inertial mode with the same north--south symmetry, these RZ Rossby modes couple with the CZ and develop substantial flows in the CZ remarkably similar to those of the CZ inertial modes.
We refer to such modes as CZ--RZ mixed modes.
We remark that these couplings would have been very different if one used uniform rotation throughout the domain, as the frequency of the RZ Rossby modes is entirely determined by the rotation rate in the RZ (see Appendix~\ref{Appendix:Uniform-rot}). 
\citet{Blume2024ApJ} discusses two separate Rossby wave cavities in the stably stratified RZ and adiabatic CZ. Our studies suggest that the separation of Rossby-mode frequencies in the CZ and RZ is primarily due to differential rotation. Furthermore, CZ Rossby modes cannot mix with sectoral Rossby modes in the RZ for $m>2$ due to their different frequencies. Nevertheless, the RZ Rossby modes can mix with other inertial modes in the CZ under the concurrent conditions identified in our work.

\begin{figure*}
    \centering
    \includegraphics[width=0.95\linewidth]{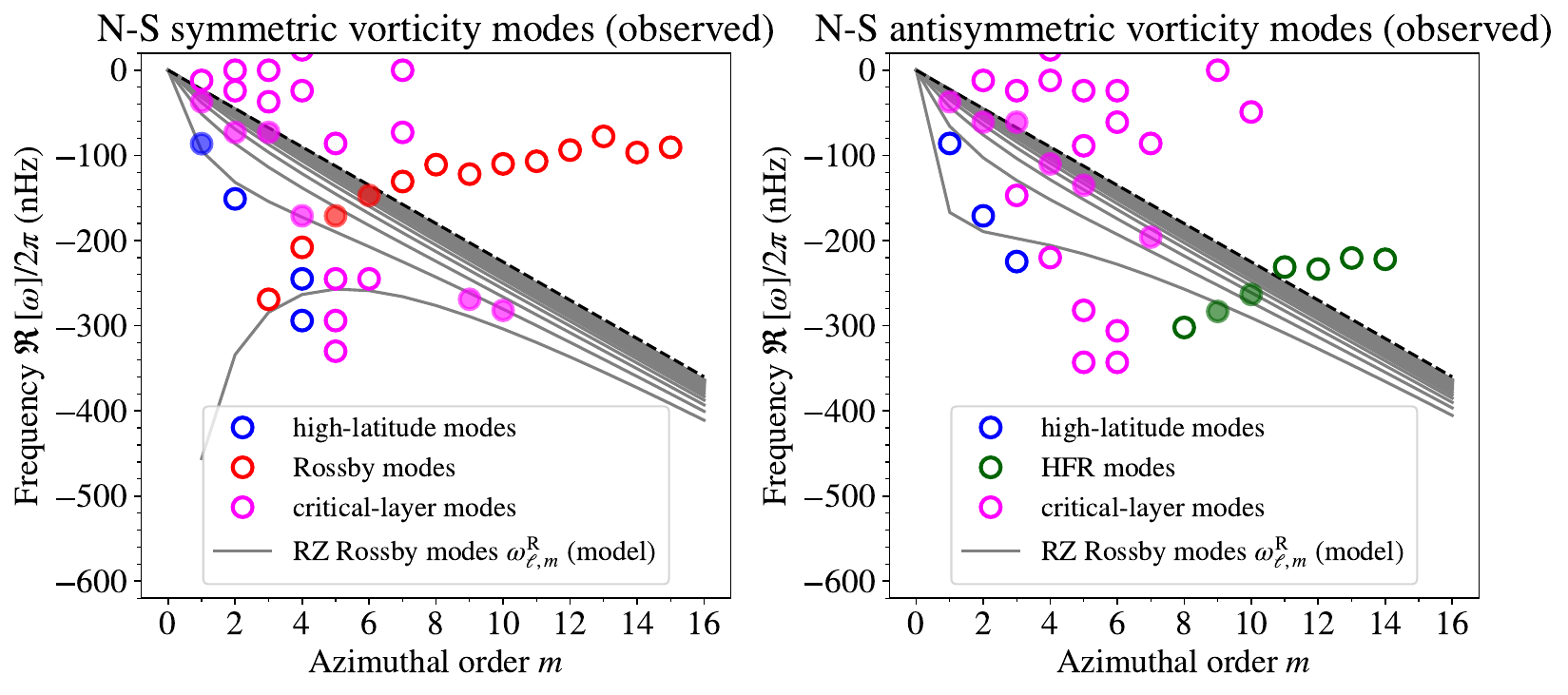}
    \caption{Frequencies of the observed solar inertial modes (open circles; from \citealt{Loptien2018NatAs, Gizon2021A&A, Hanson2022NatAs}) overplotted with the dispersion relations of the RZ Rossby modes (solid grey; given by Eq.~\ref{eq:Rossby_dispersion_RZ}).
    The left and right panels correspond to modes with north--south symmetric and antisymmetric vorticity, respectively. The observed inertial modes are high-latitude (blue), Rossby (red), HFR (green), and critical-layer (magenta) modes. Observed modes whose frequencies are within $10\,\mathrm{nHz}$ of an RZ Rossby mode frequency with the same symmetry ($|\Delta\omega|\lesssim 10\,\mathrm{nHz}$) are shown with a translucent fill.
    }
    \label{fig:11}
\end{figure*}

\subsection{Detectability of CZ--RZ mixed modes}
One may ask if any of the CZ--RZ mixed inertial modes can be observed, as many of these modes have comparable growth rates to the CZ inertial modes (see Fig.~\ref{fig:10}).
The surface eigenfunctions and frequencies of many CZ inertial modes and their nearest CZ--RZ mixed modes are likely too similar to be distinguished using surface observations alone. 
Figure~\ref{fig:11} shows the frequencies of the observed solar inertial modes. Many of the observed modes have frequencies close to those of the RZ Rossby modes, which share the same north--south symmetry. This allows them to couple with the RZ Rossby modes. It may be that the marked modes with close frequencies are weakly mixed with the RZ Rossby modes. Many such modes include critical-layer modes, which could plausibly be extensions of some RZ Rossby modes with high $n_{\rm RZ}$. 

A model for the excitation and damping of these modes would certainly help determine whether some of the mixed modes can reach observable amplitudes. For the linearly stable modes, stochastic excitation by convection is likely the dominant mechanism \citep{Philidet2023A&A}. 
However, as reported in Table~\ref{tab:1} (see also Appendix~\ref{Appendix:mode-mass}), the mixed modes are shown to possess a mode mass \citep[see e.g.][]{Christensen-Dalsgaard2002RvMP, Baudin2005A&A} much higher than that of the CZ inertial modes.
This makes it energetically difficult to excite such mixed modes by turbulent convection in the CZ.
For linearly unstable modes, dedicated non-linear simulations will be required to reliably estimate saturation amplitudes \citep{Bekki2024SciA}. Latitudinal shear instability in the tachocline may excite some RZ Rossby modes \citep{Garaud2001MNRAS}. However, we do not find such linear instabilities in our study.

\subsection{Implications of horizontal motions below the CZ}
We find that the CZ inertial modes have notably significant motions in the OS (see Fig.~\ref{fig:3}). The $m=1$ high-latitude mode has RMS velocity $\sim3~\rm m~s^{-1}$ in the OS, when normalized to the observed amplitude at the surface. The $m=3$ equatorial Rossby mode also has RMS velocity $\sim0.7~\rm m~s^{-1}$ in the OS, while the other modes shown in Fig.~\ref{fig:3} have RMS velocities $\lesssim 0.2~\rm m~s^{-1}$ in the OS. 
Furthermore, CZ--RZ mixed modes have higher velocities in the RZ as compared to the CZ because of less dissipation in the RZ (refer to Appendix~\ref{Appendix:dissipation}). 
The presence of these (mostly horizontal) motions in the sub-adiabatic OS may contribute significantly to horizontal turbulent diffusion in the solar tachocline \citep[see discussion in][]{Garaud2025ApJ}. 
They could also aid in the confinement of the tachocline through local dynamo effects \citep[e.g.][]{Matilsky2022ApJ}.

\begin{acknowledgements}
Author contributions: 
LG and XZ initiated this project; SM implemented all the equations in \texttt{Dedalus} and performed the computations and analysis; YB provided supervision to validate the results; all authors discussed the results; SM wrote the initial draft, and all authors contributed to the final manuscript. 
We thank P.~Dey, D.~Fournier, and R.~H.~Cameron for helpful discussions.
SM is a member of the International Max Planck Research School for Solar System Science at the University of G{\"o}ttingen. YB and LG acknowledge support from ERC Synergy Grant WHOLE SUN 810218. XZ acknowledges the financial support from the German Research Foundation (DFG) through grants 521319293, 540422505, and 550262949. 
\end{acknowledgements}

\bibliographystyle{aa}

\begin{appendix}

\section{Effects of dissipation in the RZ}\label{Appendix:dissipation}
\begin{figure}
    \centering
    \includegraphics[width=0.95\linewidth]{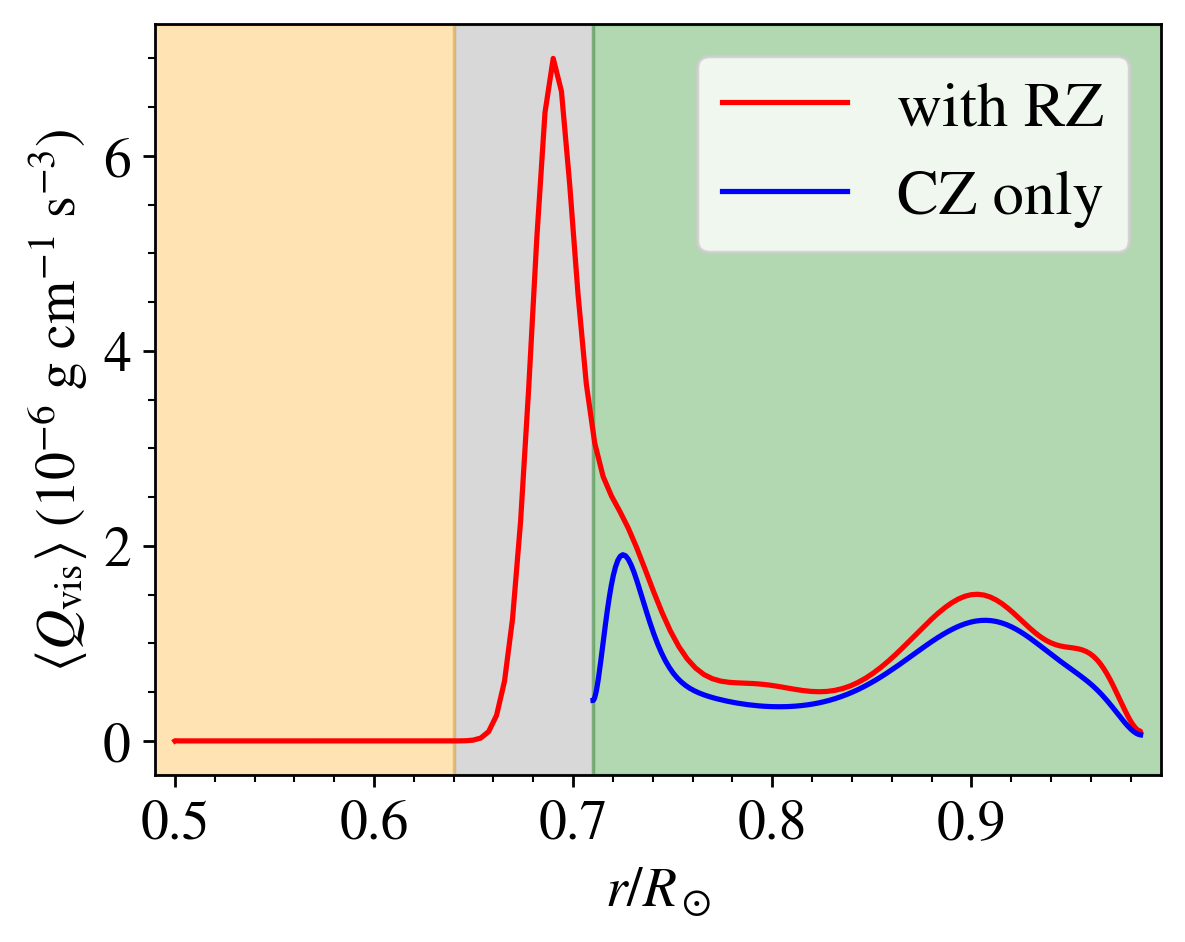}
    \caption{Radial profile of dissipation, $Q_{\rm vis}$, defined in Eq.~\eqref{eq:dissipation}, for the $m=3$ prograde columnar mode in the models with (red) and without the RZ (blue). The eigenmodes are shown in Fig.~\ref{fig:3}. They are normalized to have the same total energy density in the CZ. 
    }
    \label{fig:A1}
\end{figure}
In this appendix, we examine the effects of dissipation in the RZ.
For many of the CZ--RZ mixed modes presented in the paper, the velocity amplitude tends to be larger in the RZ than in the CZ. 
We provide a physical explanation for this behaviour. We also illustrate why the CZ inertial modes are more damped when the RZ is included. 

The equation for the kinetic energy density $e_{\rm kin}=(\rho_0/2)|\vec{u}|^2$ is obtained by taking a dot product of Eq.~\eqref{eq:1} with $\vec{u}$. Neglecting the advection terms, this yields
\begin{equation}
    \frac{\partial e_{\rm kin}}{\partial t} \approx -\vec{\nabla}\cdot(p_1\vec{u} + \nu\rho_0 \vec{u}\cdot \tens{S})-\rho_1 g u_r - Q_{\rm vis},
\end{equation}
where $Q_{\rm vis}$ denotes the viscous dissipation of kinetic energy:
\begin{equation}\label{eq:dissipation}
    Q_{\rm vis} = \frac{\rho_0}{2}\nu\  \tens{S} : \tens{S} .
\end{equation}

By taking a volume integral under appropriate radial boundary conditions, we obtain the equation for the total kinetic energy of the toroidal Rossby modes (with $u_r \approx 0$) as
\begin{equation}
    \frac{\partial}{\partial t} \int_V e_{\rm kin} dV \approx - \int_V Q_{\rm vis} dV.
\end{equation}
Therefore, the mode damping rate $\gamma = - \Im[\omega]$ can be expressed as
\begin{equation}
   2\gamma \approx \dfrac{\int_V Q_{\rm vis} dV}{\int_V e_{\rm kin} dV}.
\end{equation}
The order-of-magnitude estimate $\tens{S} : \tens{S}  \approx k^2|\vec{u}|^2$, where $k$ is the local wavenumber, implies that the damping $\gamma$ increases with the number of radial nodes.

The mode damping rate is given as the radial average of $\nu k^2$ weighted by the kinetic energy density $e_{\rm kin}$. 
In general, there is a tendency for eigenmodes to distribute more (less) kinetic energy into a region where the dissipation $\nu k^2$ is small (large) in order to minimize the mode damping rate $\gamma$.
The RZ, having a lower $\nu$, tends to have higher kinetic energy for CZ--RZ mixed modes with a low number of radial nodes.
This further helps explain the increase in motions in the CZ for CZ--RZ mixed modes as the number of radial nodes in the RZ increases (because $k$ is increased). Thereby, the RZ Rossby modes also become more damped with the increase in $n_{\rm RZ}$ (see Fig.~\ref{fig:6}).

Figures~\ref{fig:3} and~\ref{fig:4} demonstrate that the CZ inertial modes are significantly damped by the inclusion of the RZ. We demonstrate that this is caused by dissipation arising from the decay of modes in OS. Figure~\ref{fig:A1} shows the spherically averaged radial profile of dissipation $Q_{\rm vis}$ for the $m=3$ prograde columnar mode, which is the most affected mode among the CZ inertial modes whose eigenfunctions are presented in Fig.~\ref{fig:3}. The radial profile of $Q_{\rm vis}$ demonstrates that the dissipation is much higher in OS as compared to the whole of the CZ. This is because of the sharp decay of the modes in OS, where the viscosity is finite. The generated dissipation significantly damps the CZ inertial modes when the RZ is included (see Fig.~\ref{fig:4}). 

\FloatBarrier
\section{Case of uniform rotation}\label{Appendix:Uniform-rot}

\begin{figure*}
    \centering
    \includegraphics[width=0.88\linewidth]{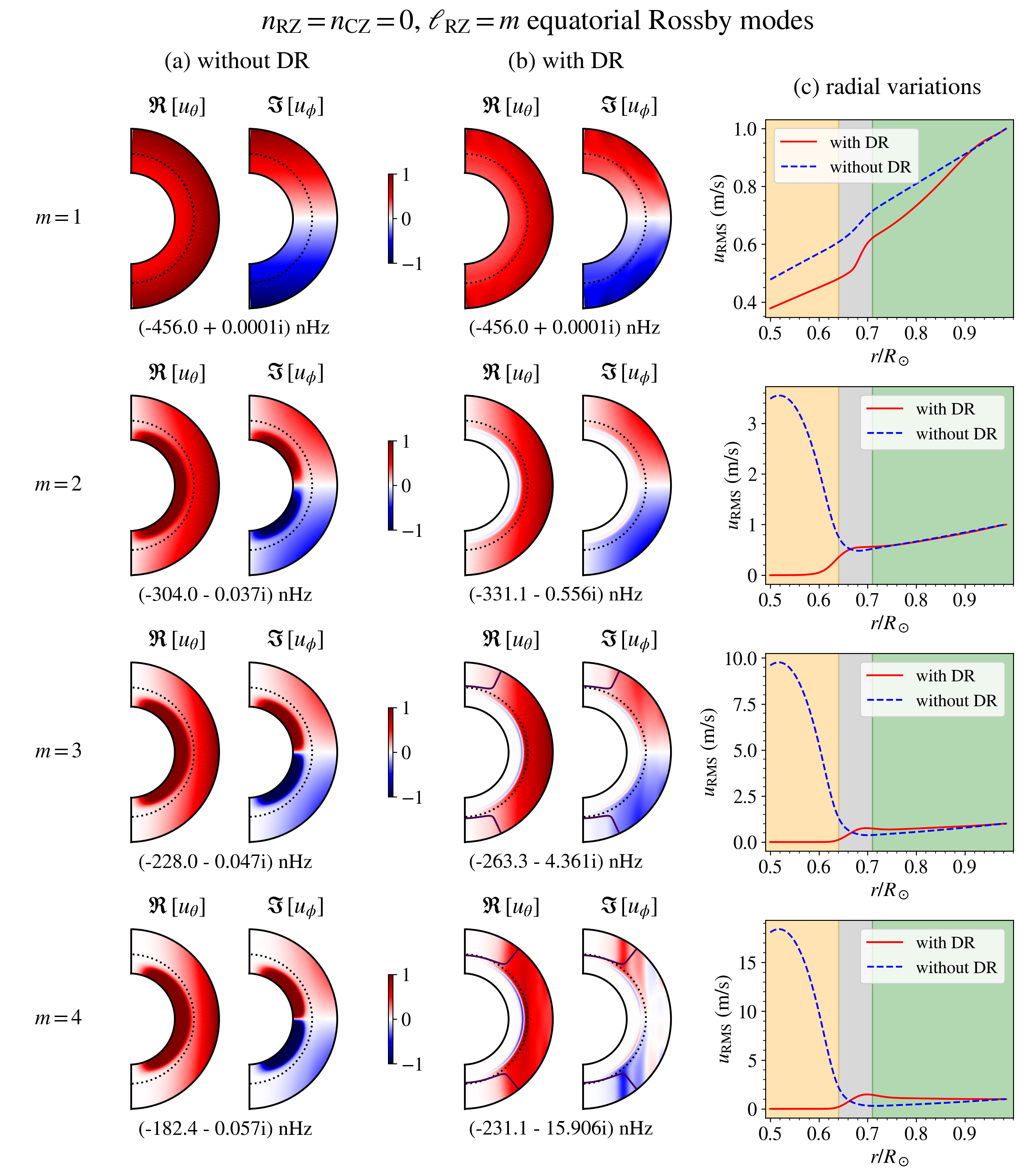}
    \caption{Analysis of the effects of differential rotation on sectoral Rossby modes with no nodes in radius ($n_{\rm RZ}=n_{\rm CZ} =0, \, \ell _{\rm RZ} = m $).
    (a) Meridional cross-sections of $u_\theta$ and $u_\phi$ of the Rossby modes computed under uniform rotation for different azimuthal orders, $m$. The longitudes corresponding to the real and imaginary components are designated in the same way as in Fig.~\ref{fig:3}. The frequencies and the growth rates measured in the Carrington frame are noted below the eigenfunctions. All eigenfunctions are normalized to have a maximum $u_\theta$ of $1\, \rm m~s^{-1}$ at the surface. The dotted lines indicate the position of the CZ base. (b) Same as panel a but under the simplified differential rotation given by Eq.~\eqref{eq:simpleDR}, shown in Fig.~\ref{fig:1}. 
    (c) Radial profiles of the RMS velocity of the modes computed with and without differential rotation. The colours of the shaded regions have the same meaning as in Fig.~\ref{fig:2}.}
    \label{fig:B1}
\end{figure*}

In Sect.~\ref{sec:3.2}, we show that the Rossby modes in the CZ and RZ are separated in frequency because of differential rotation. 
This appendix discusses the case of uniform rotation throughout the Sun, i.e. $\Omega(r,\theta) = \Omega_0$, while maintaining the radial profiles of all other background quantities.
We focus here on the sectoral ($\ell=m$) Rossby modes, because the non-sectoral modes cannot exist in the CZ.
Figure~\ref{fig:B1} compares the eigenfunctions of the sectoral modes with $1\leq m \leq 4$ under uniform rotation and those under differential rotation.
In the case of uniform rotation, the mode frequencies are given by $\omega_{m}=-2\Omega_0/(m+1)$ in both the CZ and RZ, so the eigenmodes extend globally.
In other words, there is essentially no cavity separating the Rossby modes in the RZ from those in the CZ.
We note, however, that the underlying force balance maintaining their toroidal nature is different in these two regions: The radial pressure gradient force is balanced by the Coriolis force in the CZ but by buoyancy in the RZ (see Fig.~\ref{fig:7}).
Consequently, the radial eigenfunctions do not show the well-known $r^m$ dependence (predicted assuming the geostrophic balance) inside the RZ.
Figure~\ref{fig:B1} also shows that the mode tends to have a larger velocity amplitude in the RZ than in the CZ as $m$ increases.
This is likely because, when the local wavenumber $k$ increases with $m$, the eigenmode tends to concentrate more of its kinetic energy in the less viscous RZ in order to minimize the damping rate (see Appendix~\ref{Appendix:dissipation}).

In the case of differential rotation in the CZ, on the other hand, the situation is markedly different.
The sectoral Rossby modes in the RZ and the equatorial Rossby modes in the CZ (distorted from the original sectoral spherical harmonics by critical layers) have different frequencies, and thus they cannot form a single global eigenmode but instead exist only as isolated, distinct modes.
A similar result has been found by \citet{Blume2024ApJ}, who reported that the CZ and RZ host separate wave cavities for Rossby modes.
We note that the only exception is the $\ell=m=1$ mode, in which the CZ and RZ are strongly coupled. 
This particular mode, known as the spin-over mode \citep[e.g.][]{Kerswell1993GApFD}, has a fixed frequency of $-\Omega_0$ and exhibits global motions with quasi-uniform vorticity in a direction perpendicular to the rotational axis \citep[][]{Greenspan1968Book}.
Owing to its special flow structure, this mode experiences negligible viscous dissipation.
Consequently, unlike the other modes discussed above, the spin-over mode does not need to redistribute its kinetic energy in response to local viscosity, allowing it to have nearly the same eigenfunction in the CZ and the RZ. This mode, however, is difficult to detect because its frequency is nearly zero in an inertial frame ($-31.7$~nHz in the Earth’s frame), placing it in a spectral region often dominated by systematics and signals associated with Earth’s revolution around the Sun.

\FloatBarrier
\section{Analysis of HFR modes}\label{Appendix:HFR}
\begin{figure}
    \centering
    \includegraphics[width=0.95\linewidth]{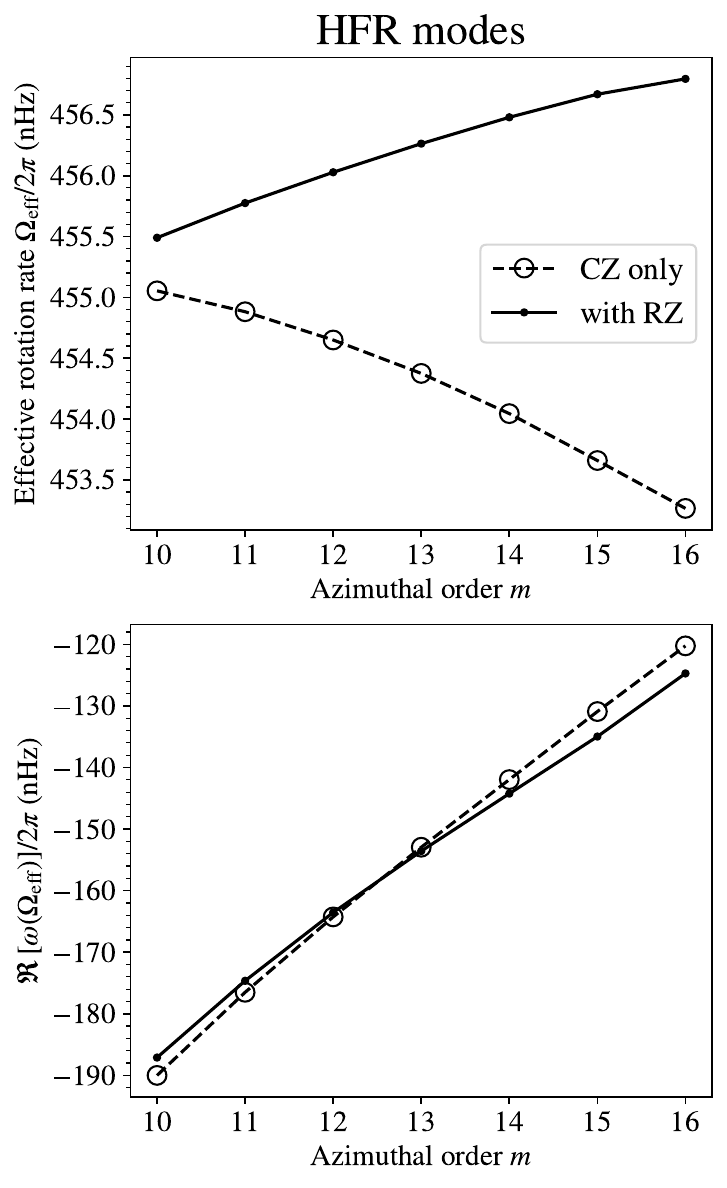}
    \caption{\textit{Top}: Effective rotation rate ($\Omega_{\rm eff}$), given by Eq.~\eqref{eq:Ome_eff}, for the HFR modes in the models that include the RZ (solid line with points) and exclude the RZ (dashed line with open circles). \textit{Bottom}: Frequency, $\omega(\Omega_{\rm eff})$, of the HFR modes in the frame rotating at $\Omega_{\rm eff}$, calculated using Eq.~\eqref{eq:freq_shift}, with the same notation. 
    }
    \label{fig:C1}
\end{figure}
In this appendix, we investigate the origin of the significant difference in the dispersion relations of HFR modes at $m\geq 12$ between the models that include and exclude the RZ (see Fig.~\ref{fig:4}). We analyse the effective rotation rate of HFR modes with $10 \leq m \leq 16$ because the dispersion relations begin to diverge at $m=10$. Although the differences are small for $m<12$, they become large, exceeding 20 nHz, only for $m>12$. 
The effective rotation rate of the mode can be measured by
\begin{equation}\label{eq:Ome_eff}
    \Omega_{\rm eff} = \frac{\int_V \Omega(r,\theta) e_{\rm kin}\, dV}{\int_V e_{\rm kin}\, dV}.
\end{equation}
 In the Carrington frame, a change in $\Omega_{\rm eff}$ causes a Doppler shift in the frequency given by
\begin{equation}\label{eq:freq_shift}
    \omega = \omega(\Omega_{\rm eff}) + m(\Omega_{\rm eff} -\Omega_0),
\end{equation}
where $\omega$ is the frequency in the Carrington frame and $\omega(\Omega_{\rm eff})$ is that in the frame rotating at $\Omega_{\rm eff}$. At high $m$, small differences in $\Omega_{\rm eff}$ lead to significant differences in $\omega$.

The top panel of Fig.~\ref{fig:C1} shows that the effective rotation rates $\Omega_{\rm eff}$ of the HFR modes diverge between the models with and without the RZ as $m$ increases.
The bottom panel of Fig.~\ref{fig:C1} compares the frequencies of the HFR modes in the frame rotating at $\Omega_{\rm eff}$, as obtained from Eq.~\eqref{eq:freq_shift}. It indicates that $\omega(\Omega_{\rm eff})$ is indeed very similar for the HFR modes in the two models. 
The differences in $\omega(\Omega_{\rm eff})$ between the two models are of the same order as those in $\Omega_{\rm eff}$. 
Therefore, it is demonstrated that the large deviation in the HFR mode frequencies seen in Fig.~\ref{fig:4} is simply caused by the difference in the effective rotation rates.

\FloatBarrier
\section{Mode mass of the mixed modes}\label{Appendix:mode-mass}
\begin{figure*}
    \centering
    \includegraphics[width=0.95\linewidth]{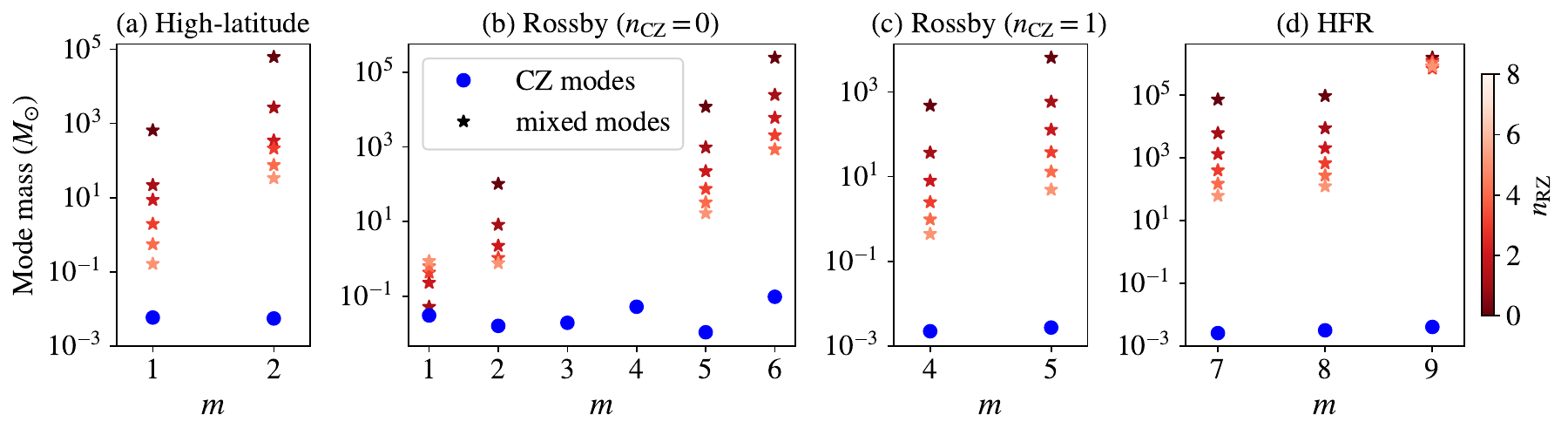}
    \caption{Mode mass (Eq.~\ref{eq:mode-mass}) of the mixed modes (red stars) and of their nearest inertial modes with motions predominantly in the CZ (blue dots). The 11 mixed modes marked in Fig.~\ref{fig:8} are shown. The colour of the star symbols denotes the number of radial nodes ($n_{\rm RZ}$) in the region $0.5R_\odot\leq r\leq 0.71 R_\odot$. For simplicity, we show only the mixed modes with $n_{\rm RZ}\leq 5$.
    }
    \label{fig:D1}
\end{figure*}
In this appendix, we present the mode masses of the computed eigenmodes \citep[see e.g.][]{Baudin2005A&A}. 
The mode mass of an eigenmode is defined as
\begin{equation}\label{eq:mode-mass}
     \mathcal{M}=\frac{1}{u^2_{\rm RMS}(r_s)} \int_V \rho_0 |\vec{u}|^2 dV,
\end{equation}
where $u^2_{\rm RMS}(r_s)$ is the observed mode amplitude at the surface $r_{\rm s}$. 
Note that the mode mass is independent of the choice of mode normalization. 
Modes with higher mode masses are energetically more difficult to excite \citep{Samadi2001A&A}.

Figure~\ref{fig:D1} presents the mode masses of the CZ--RZ mixed modes and their corresponding CZ inertial modes. 
It is shown that the mixed modes possess higher mode masses compared to the CZ inertial modes, and the mode mass tends to decrease as $n_{\rm RZ}$ increases.
We note that, owing to very large velocity amplitudes in the RZ, mode masses of some mixed modes can exceed the solar mass $M_\odot$ by several orders of magnitude.
These modes are very unlikely to be excited stochastically by turbulent convection in the CZ.
Only mixed modes with sufficiently large $n_{\rm RZ}$ can have mode masses close to those of the CZ inertial modes. 
On the other hand, our study also reveals that the growth rate of the mixed modes decreases as the number of radial nodes, $n_{\rm RZ}$, increases (see Fig.~\ref{fig:6}). 
These competing factors may determine the allowed range of $n_{\rm RZ}$ for the mixed modes that can be excited to observable amplitudes at the surface.

\FloatBarrier
\section{Convergence in resolution}\label{Appendix:resolution}
\begin{figure*}[htbp]
    \centering
\includegraphics[width=0.95\linewidth]{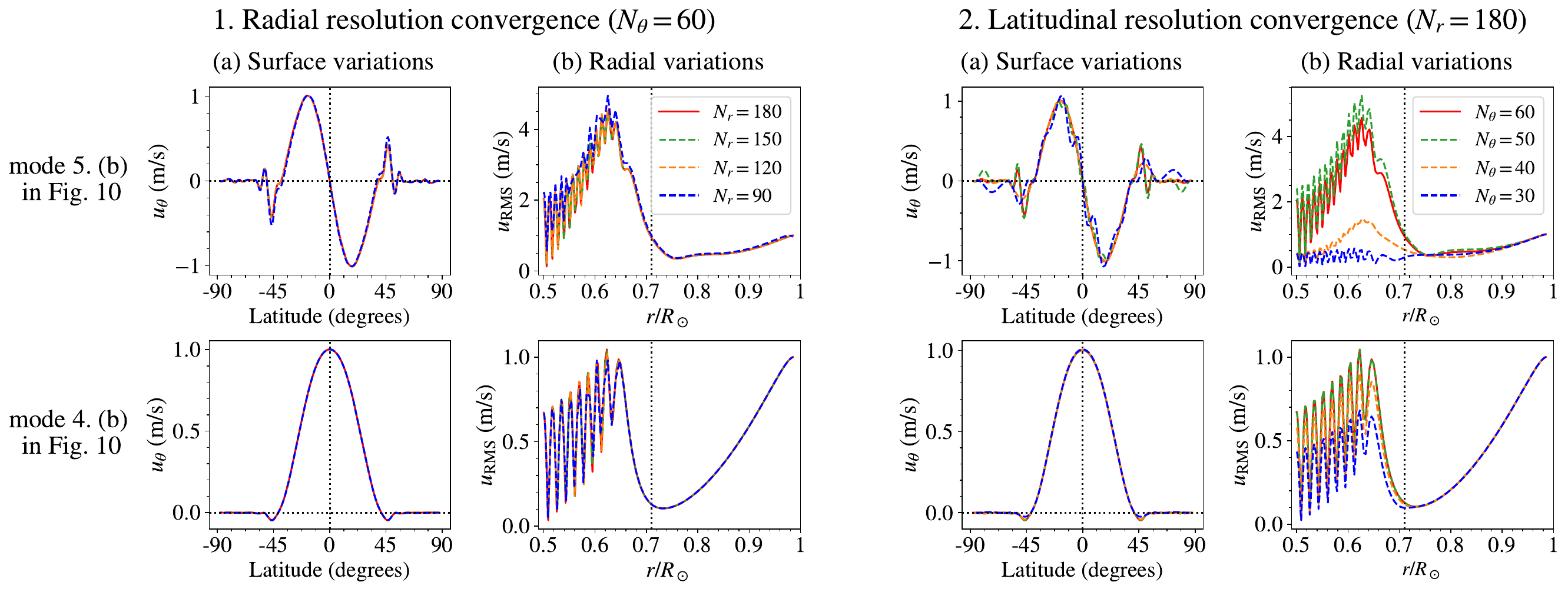}
    \caption{Convergence with respect to grid resolution for two representative CZ--RZ mixed modes with strong spatial variations.
    The top and bottom rows show the results for the modes labelled 5. (b) and 4. (b) in Fig.~\ref{fig:10}, respectively.
    Panel $1$: Mode structures for various radial resolutions ($N_r$) with fixed latitudinal resolution $N_\theta=60$. 
    Panel $2$: Mode structures for various $N_\theta$ with fixed $N_r=180$.
    Surface and radial variations of the modes are represented in panels (a) and (b), respectively, in the same way as in Fig.~\ref{fig:10}. 
    }
    \label{fig:E1}
\end{figure*}

\begin{table}[!h]
\caption{Frequency convergence with radial resolution, $N_r$.}
    \centering
    \begin{tabular}{ccc}
    \hline \hline
       $N_r$  & Mode 5. (b)  & Mode 4. (b)  \\ \hline
        180 & $-240.068-25.593$i & $-136.740-11.067$i    \\
        150 & $-240.070-25.524$i & $-136.743-11.070$i   \\
         120& $-240.073-25.581$i & $-136.739-11.070$i \\ 
         90 & $-240.324-25.815$i & $-136.736-11.077$i \\\hline
    \end{tabular}
    \tablefoot{The two modes are labelled as modes 5. (b) and 4. (b) in Fig.~\ref{fig:10}. The complex eigenfrequencies, $\omega/2\pi$, are reported in units of nHz. In all cases, the latitudinal resolution is fixed at $N_\theta=60$.}
    \label{tab:2}
\end{table}

\begin{table}[!h]
\caption{Frequency convergence with latitudinal resolution, $N_\theta$.}
    \centering
    \begin{tabular}{ccc}
    \hline \hline
       $N_\theta$  & Mode 5. (b)  & Mode 4. (b)   \\ \hline
        60 & $-240.068-25.593$i & $-136.740-11.067$i    \\
        50 & $-240.112-25.618$i & $-136.740-11.067$i   \\
         40& $-239.401-24.789$i & $-136.740-11.064$i \\ 
         30 & $-241.352-25.832$i & $-136.765-11.083$i \\\hline
    \end{tabular}
    \tablefoot{Same convergence check as in Table~\ref{tab:2}, but with respect to $N_\theta$. The radial resolution is fixed at $N_r=180$.}
    \label{tab:3}
\end{table}

In this appendix, we present supplementary computations to demonstrate that both the complex eigenfrequencies and the eigenfunctions of the modes presented in this study are well resolved. 
To this end, we examine the convergence with resolution for two selected modes: the CZ--RZ mixed HFR mode (corresponding to the mode 5. (b) in Fig.~\ref{fig:10}), which has the largest radial variations (with $n_{\rm RZ}=13$), and the CZ--RZ mixed $n_{\rm CZ}=1$ Rossby mode (corresponding to the mode 4. (b) in Fig.~\ref{fig:10}), which has the largest latitudinal variations (with $\ell_{\rm RZ}-m =8$).
Figure~\ref{fig:E1} shows that the eigenfunctions of these modes are almost unaltered for $N_r\geq120$ and $N_\theta\geq 50$, exhibiting radial and latitudinal resolution convergence for both modes.
Tables~\ref{tab:2} and \ref{tab:3} further show that the eigenfrequencies are also well converged within a difference of $0.1$ nHz for $N_r\geq120$ and $N_\theta\geq 50$. 
This ensures that all other modes presented in this study, which have lower radial and latitudinal variations, are also converged.

\end{appendix}

\end{document}